\documentclass[final, 3p]{elsarticle}
\usepackage{graphicx}
\usepackage{amsmath}
\usepackage{amsfonts}
\usepackage{mathrsfs}
\usepackage[OT1]{fontenc} 
\usepackage{enumerate}
\usepackage{amsthm}
\usepackage{tabularx}
\usepackage{arydshln}

\usepackage[colorlinks=true,linkcolor=blue,citecolor=blue]{hyperref}
\usepackage{mwe}
\biboptions{authoryear, sort}
\usepackage{subfig}
\usepackage{float}
\usepackage{graphicx}
\usepackage{textcomp}
\usepackage{xcolor}
\usepackage{footnote}
\usepackage{makecell}
\usepackage{graphicx}
\usepackage{amsmath}
\usepackage{amsfonts}
\usepackage{mathrsfs}
\usepackage[OT1]{fontenc} 
\usepackage{enumerate}
\usepackage{amsthm}
\usepackage{tabularx}
\usepackage{arydshln}
\usepackage{fixfoot}
\colorlet{re}{black}
\colorlet{r2}{black} 
\colorlet{r3}{black} 
\usepackage{colortbl}
\usepackage{tikz}

\usepackage{subfig}
\usepackage{textcomp}
\usepackage{xcolor}
\usepackage{footnote}
\usepackage{makecell}
\usepackage[draft]{todonotes}

\usepackage{algorithm}
\usepackage{algpseudocode}
\usepackage[inline]{enumitem}
\usepackage{mathtools}

\begin{document} 

\begin{frontmatter}

\title{Influence of CAV Clustering Strategies on Mixed Traffic Flow Characteristics: An Analysis of Vehicle Trajectory Data}

\author[1]{Zijia Zhong\corref{cor1}}
\address[1]{Department of Civil and Environmental Engineering, University of Delaware, United States}
\cortext[cor1]{Corresponding author: Zijia Zhong. The research was conducted at the University of Delaware; the author is now with the Center for Integrated Mobility Sciences at the National Renewable Energy Laboratory.
}
\ead{zijia.zhong@nrel.gov}

\author[1]{Earl E. Lee}
\ead{elee@udel.edu}

\author[1]{Mark Nejad}
\ead{nejad@udel.edu}

\author[2]{Joyoung Lee}
\address[2]{John A. Reif, Jr. Department of Civil and Environmental Engineering, New Jersey Institute of Technology, United States}
\ead{jo.y.lee@njit.edu}

\textcolor{red}{Please cite this article as: Z. Zhong, E. E.Lee, M. Nejad and J. Lee, ``Influence of CAV clustering strategies on mixed traffic flow characteristics: an analysis of vehicle trajectory data,'' Transp. Res. Part C: Emerging Technologies, vol.115, pp.102611, 2020, doi:10.1016/j.trc.2020.102611}

\begin{abstract}
\textcolor{r2}{Being one of the most promising applications enabled by connected and automated vehicles (CAV) technology,  Cooperative Adaptive Cruise Control (CACC) is expected to be deployed in the near term on public roads.} Thus far, the majority of the CACC studies have been focusing on the overall network performance with limited insights on the potential impacts of CAVs on human-driven vehicles (HVs).
This paper aims to quantify such impacts by studying the high-resolution vehicle trajectory data that are obtained from microscopic simulation. Two platoon clustering strategies for CACC- an ad hoc coordination strategy and a local coordination strategy-are implemented. 
Results show that the local coordination outperforms the ad hoc coordination across all tested market penetration rates (MPRs) in terms of network throughput and productivity. According to the two-sample Kolmogorov-\textcolor{re}{Smirnov} test, however, the distributions of the hard braking events (as a potential safety impact) for HVs change significantly under local coordination strategy. For both of the clustering strategy, CAVs increase the average lane change frequency for HVs. The break-even point for average lane change frequency between the two strategies is observed at 30\% MPR, which decreases from 5.42 to 5.38 per vehicle.  The average lane change frequency following a monotonically increasing pattern in response to MPR, and it reaches the highest 5.48 per vehicle at 40\% MPR. Lastly, the interaction state of the car-following model for HVs is analyzed. It is revealed that the composition of the interaction state could be influenced by CAVs as well. One of the apparent trends is that the time spent on approaching state declines with the increasing presence of CAVs. 
\end{abstract}
  

\end{frontmatter}


\section{Introduction}
Cooperative adaptive cruise control (CACC) enables closely-coupled vehicular platoons by extra layers of communication and automation. Being one of the most-studied applications of the connected and automated vehicle (CAV) technology,  CACC is expected to drastically increase mobility and decrease emissions, while providing a safer and more convenient way for vehicle occupants. 
Studying CACC under mixed traffic conditions in anticipation of its near-term deployment has gained a significant amount of attention \citep{ zhao2018aplatoon, Wang2019stability, Amirgholy2020traffic, Ghiasi2019mixed, Saha2019preferred}. \textcolor{re}{The near-term deployment of CACC likely entails the following conditions:
\begin{enumerate*}[label=\roman*)]
\item relatively low CACC market penetration (e.g., less than 40\%),
\item the substantial presence of non-equipped vehicle without communication capability (i.e., human-driven vehicles (HVs)),
\item legacy roadway infrastructure that is not optimized for CACC operation (e.g., mixed-use travel lane, lack of roadside units).
\end{enumerate*}
Local coordination allows CAV vehicles within a certain distance to form a string of CAVs (or vehicular platoon). Such a strategy is appealing to minimize CACC degradation \citep{Wang2019stability}, where a CAV operates under ACC mode. 
}

Thus far, the CACC evaluation has been focusing on the benefits that CAV could potentially bring to the transportation networks. However, the potential impacts of platoon clustering strategies on non-equipped vehicles (i.e., HVs) have been seldom studied.
\textcolor{re}{
Empirical trajectory data revealed that lane change caused by merging and diverging vehicle creates most motorway turbulence \citep{vanBeinum2018Driving}. Therefore, CACC may influence the characteristics of the HV flow when a local coordination strategy is employed. More specifically, the expected potential impacts of CACC systems on HVs may include:
\begin{enumerate*}[label=\roman*)]
\item additional weaving during CACC clustering, 
\item lane changes by HVs induced by the aforementioned CACC clustering,
\item increase collision risks, and
\item lane blockages by CACC platoons.
\end{enumerate*}
}

\textcolor{re}{ In this study, the traffic flow characteristics of the HVs, instead of the CAVs, are examined when ad hoc coordination and local coordination strategies for CACC is deployed. The interactions between HVs and CAVs are analyzed based on the high-resolution vehicle trajectory data that are extracted from microscopic simulations. The interaction state among HVs is also studied to derive insights. The proposed method is also suitable for extracting the driving behavioral data for future field deployment and modeling heterogeneous traffic flow that is consisted of HVs and CAVs. 
}

The organization of the remainder of the paper is as follows. Relevant literature regarding studies of CACC in mixed traffic is reviewed in Section \ref{sect:Literature}, followed by the evaluation method in Section \ref{sect:framework}. The simulation results are discussed in Section \ref{sect:result}. Lastly, findings and recommendations are presented in Section \ref{sect:conclusion}.

\section{Relevant Research}
\label{sect:Literature}

CACC can positively improve traffic performance with a sufficient presence, which is commonly expressed as the market penetration rate (MPR). \textcolor{re}{MPR is the ratio of a product or service that is being used by customers and the total estimated demand. 
As the CACC technology advances, we have seen an increasing amount of field experiments for CACC \citep{MILANES2014285,xu2019design, aarts2016european, Ma2019anEco, chang2018connected}. However, the majority of them have a limited scale and focus on the hardware aspect of the vehicle. It is difficult to extrapolate the testing results at the traffic flow level. Thus far, analytical and simulation approaches remain the two primary methods for evaluating the traffic impact of CACC.
Analytical approaches mainly based on the macroscopic traffic flow model.  It may not faithfully capture the nonlinear phenomena in traffic flow (e.g., lane drop), and it only identifies the impact of a limited number of factors.  In comparison, microscopic traffic simulation is capable of capturing more complex traffic phenomena and has been widely adopted.
}

The effectiveness of CACC depends on three major factors:
\begin{enumerate*}[label=\roman*)]
\item achievable headway, 
\item car-following model, and
\item market penetration rate (MPR).
\end{enumerate*}
\textcolor{re}{
The reduced time headway and following distance have been recognized as primary benefits of CACC. CACC string can operate with an 0.6 s intra-platoon headway, compared to the 1.4 s or longer time gap for humans \citep{nowakowski2010cooperative}. Some CACC studies aimed to estimate the achievable roadway capacity with simple network topology (e.g., single-lane freeway).} \cite{van2006impact} assessed the impact of CACC on freeway traffic flow on a 6-km, one-lane freeway with ramps distributed with a 1.6-km interval. They found the roadway capacity reached 4,250 vehicles per hour (vph) per lane with full CACC penetration. Using the same CACC model, \cite{shladover2012impacts}  found the lane capacity of 3,600 vph at 90\% MPR on a  one-lane freeway. \cite{talebpour2016influence} proposed an analytical framework for modeling the longitudinal interaction among regular, connected, and autonomous vehicles with the focus on investigating the string stability of the heterogeneous flow. The simulation was conducted on a hypothetical one-lane highway, and mixed traffic flow performance was measured at an aggregated level. 
However, the single-lane freeway cases are prone to overestimate the mobility benefits brought by CACC by excluding lateral interaction or lane changing among the multi-lane segment on the mainline. Hence evaluation of multi-lane scenarios provide more realistic estimation, and they can be found in \cite{Arnaout2014, lee2014mobility, songchitruksa2016incorporating}

\textcolor{re}{
The heterogeneous traffic flow attributed to partial market penetration has been examined in recent years. \cite{wang2017Comparing} proposed a second-order traffic flow model for mixed traffic streams with HVs and AVs.
\cite{Xiao2019unravelling} studied the traffic flow characteristics using a multi-regime car-following model that is capable of switching among human-driven, ACC, and CACC modes. 
The Lane Change Model with Relaxation and Synchronization \citep{Schakel2012} was adopted by Xiao et al., but no clustering strategy for CACC was implemented in the study conducted. 
In mixed traffic flow, the degradation of CACC occurs when a CACC vehicle follows a non-equipped vehicle that does not has communication capability. Under this circumstance, a CACC vehicle reverts to traditional ACC mode with a more conservative target headway. The negative effect of ACC on the system capacity is due to the higher time gap of ACC \citep{Calvert2017will}. \cite{Wang2019stability} investigated the degradation of CACC in mixed traffic and the resulting flow stability. They found that the degraded CACC vehicles, based on current technical maturity, contributed negatively to the stability of the heterogeneous flow.
} 

\textcolor{re}{
Insofar, the studies on the CACC impact on HVs are relatively scarce in the literature. The majority of the CACC study for human factors pertains to the human-machine interface, especially for the transition of authority at a low or medium level of automation \citep{shen2017assessing, Lank2011Interaction, berghofer2018prediction, naujoks2017driving}. }
\cite{nowakowski2010cooperative, nowakowski2011cooperative} studied the acceptance of the short following time headway (ranging from 0.6 s to 1.1 s) enabled by CACC. As discovered, while all the drivers showed the willingness to accept the shorter following gaps, male participants were more likely to choose a shorter following distance.  The carry-over effect of the short headway in manual driving was exhibited even after the disengagement from the platoons in the KONVOI project \citep{casey1992changes}.
\cite{GOUY2014264} investigated the behavioral adaptation effects that potentially caused by the short headway of CACC platoon using driving simulators. Participants were instructed to drive alongside two CACC platoon configurations:
\begin{enumerate*}[label=\roman*)]
\item a 10-truck platoon with 0.3-s intra-platoon headway, and 
\item a 3-truck platoon with 1.4-s intra-platoon headway.
\end{enumerate*}
A smaller average time headway was observed when in the short headway scenario. In the first platoon scenarios, participants spent more time under a 1-s headway, which is deemed unsafe \citep{FAIRCLOUGH1997387}. 

\textcolor{re}{
The safety of the heterogeneous flow has been studied as well. \cite{Papadoulis2019evaluating} assessed the safety benefit of CAV and found 12-47\% reduction in traffic conflict at 25\% MPR. Local coordination among CAV was implemented with a threshold of 0.6 s form the vehicle upstream and downstream in the target lane. The safety implication of the driving aggressiveness of AV for the mixed traffic condition was studied by \cite{Lee2019driving}. The conflict rate, which is derived from the vehicle interactions between AV and HV at each simulation time step, was used as a safety indicator among nine levels of CAV aggressiveness. However, the performance of each vehicle group was not separated, and the efficacy of simulating AV behavior by merely adjusting the Wiedemann model is still subject to debate \citep{MILANES2014285}.
}

\textcolor{re}{
In anticipation of the long lead-time for full market penetration, operation strategies of CACC in mixed traffic conditions have also been actively investigated. Managed lane, vehicle awareness device (VAD), and local coordination are among the most popular operation strategies. The underlying goal of these three methods is to alleviate the CACC degradation.
Managed lane has a track record in incentivizing desired travel patterns or behaviors \citep{shewmake2014hybrid}. 
\cite{liu2018Modeling} presented a modeling framework for mixed traffic under various CACC operation strategies. The study revealed the quadratic relationship between pipeline capacity and MPR. 
Providing priority or exclusive access for CACC to managed lane could create concentrated segments of CACC vehicles, hence mitigating the degradation effects as demonstrated in previous studies \citep{hussain2016freeway, Ghiasi2017amixed, zhang2018operational, qom2016evaluation, liu2018Modeling, zhong2019effectiveness}.
}

\begin{figure}[h]
\begin{minipage}[h]{\linewidth}
\centering
\subfloat[ad hoc coordination]{\includegraphics[scale=0.32]{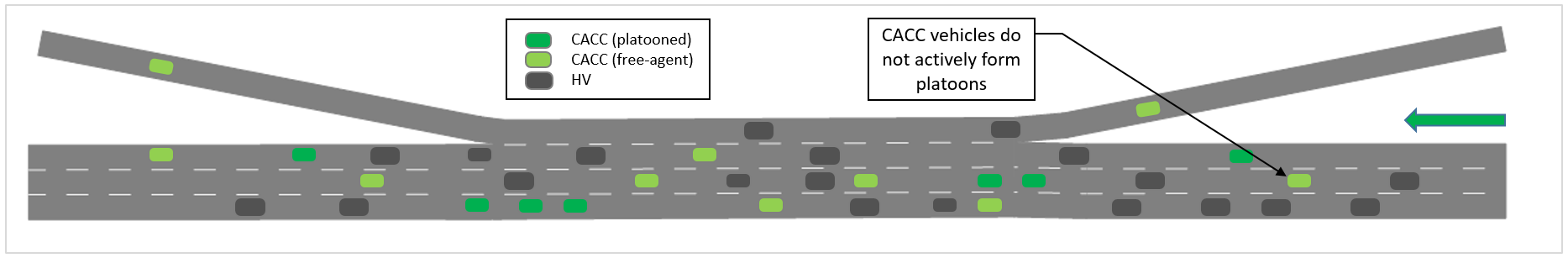}}
\end{minipage} \par
\begin{minipage}[h]{\linewidth}
\centering
\subfloat[local coordination]{\includegraphics[scale=0.32]{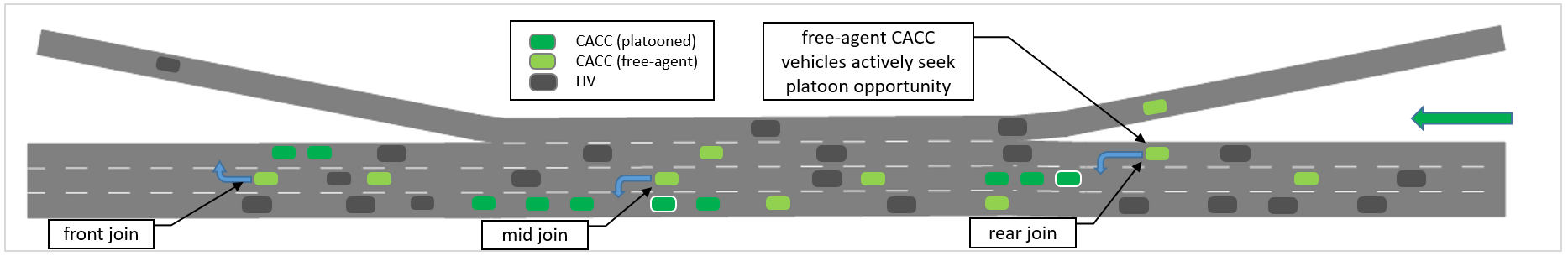}}
\end{minipage}\par
\begin{minipage}[h]{\linewidth}
\centering
\subfloat[global coordination]{\includegraphics[scale=0.32]{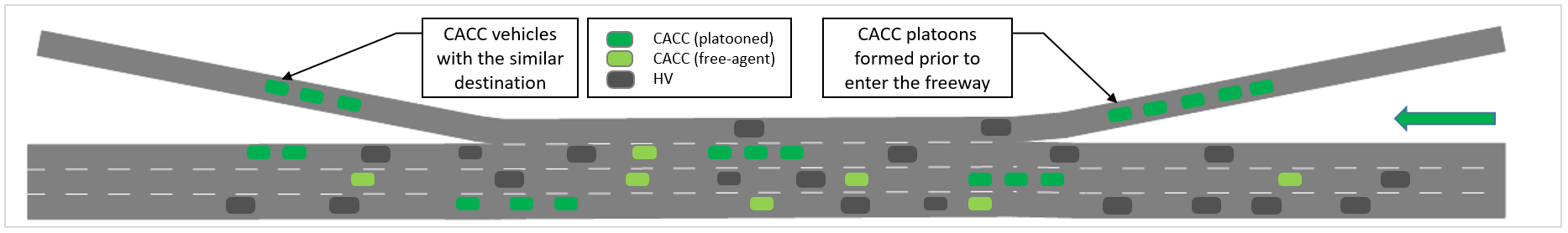}}
\end{minipage}\par
\caption{CACC clustering strategy}
\label{fig:caccCoord}
\end{figure}

\textcolor{re}{
A vehicle awareness device (VAD) is the communication device that enables a HV to transmit real-time information to the surrounding CACC vehicles. Studies found that VADs can reduce CACC degradation by enabling quicker reactions of CACC vehicles \citep{shladover2012impacts, liu2018Modeling}, and it has emerged as one of the most promising strategies.
}
The third operation strategy, which is of high relevance in this study, is the CACC coordination. As illustrated in Fig. \ref{fig:caccCoord}, there are three types of clustering strategies: ad hoc (AH) coordination, local coordination (LC), and global coordination \citep{shladover2012impacts}. In this study, only the former two clustering strategies are evaluated. The global coordination requires advance planning for the travel demand at the origin-designation level. CAVs are coordinated to enter the highway in platoons. Therefore, the impacts of global coordination probably concentrate at the merging area in the vicinity of the on-ramps, which does not fall within the scope of this study.

Ad hoc coordination assumes random arrival of CAVs and no coordination among them. Therefore, the probability of driving behind another CAV is highly correlated to MPR. Ad hoc coordination has been observed in the majority of the research due to its simplicity in implementation and the lack of the conceptual framework for CAV platoon formation.  However, the ad hoc coordination is not likely to harness the full potential of CACC, as it does not fully utilize the short intra-platoon headway enabled by the CAV technology.  Exclusive lane, or other forms of the managed lane, has been employed to aid the ad hoc coordination \citep{segata2012simulation}. 

Local coordination facilitates platoon formations, where free-agent CAVs are actively seeking clustering opportunities in their surroundings. It has been demonstrated in several notable field experiments, such as SATRE \citep{chan2016sartre}, COMPANION \citep{eilers2015companion}. The subject CAV, as well as the surrounding CAVs, can be coordinated to change their trajectories to facilitate clustering. There are four basic types of lane change:  
\begin{enumerate*}[label = \roman*)]
\item free-agent-to-free-agent  lane change, 
\item free-agent-to-platoon lane change,
\item platoon-to-free-agent lane change, and 
\item platoon-to-platoon lane change \citep{wang2017developing}. 
\end{enumerate*}
\cite{lee2014mobility} developed a local coordination scheme that allows three ways to form a platoon:
\begin{enumerate*}[label = \roman*)]
\item front-join,
\item mid-join, and 
\item rear-join.
\end{enumerate*}
\cite{songchitruksa2016incorporating} deployed a local coordination algorithm for CACC evaluation, where only rear-join to a platoon was allowed.
Improving from the coordination model of Lee et al., \cite{Zhong2017a} implemented the MIXIC (MICroscopic model for Simulation of Intelligent Cruise control) model developed by \cite{van2006impact} as a longitudinal control to study the CAV benefits for signalized arterials. The string formation and dispersion mechanism were integrated to the model of Lee et al. in an NCHRP study \citep{NAP25366}, including preferential lane logic and platoon size restriction.  A CACC control model was developed by \cite{zhong2018assessing}, where E-IDM \citep{Kesting2010} was adopted jointly with the MOBIL model \citep{Kesting2007} that prevents the lane changing of a free-agent CAV, which may be potentially disruptive to the surrounding traffic. When a potential platooning opportunity is identified via V2V communication, the CACC system estimates the impacts on the immediate vehicles based on MOBIL should the lane change be initiated.  
The Lane-change Model with Relaxation and Synchronization (LMRS) \citep{Schakel2012} and the IDM+ \citep{Schakel2010effects} were adopted jointly by \cite{Calvert2017will} for evaluating the ACC. 
The LMRS model gives a normalized strategic lane-change score by taking into consideration of route, speed gain, and lane preference. The driver is willing to accept a smaller headway and to decelerate more in LMRS for a higher desire score.

To sum up, microscopic traffic simulation has been widely adopted in assessing the benefits of CACC.
\textcolor{r3}{ Simulation study, however, is not without its caveats, and calibration of the simulation network is one of the them.
Besides a proven microscopic behavioral model, the simulation network has to be first calibrated using real-world traffic data to ensure the validity of the simulation results.  Not only the global macroscopic traffic flow characteristics (e.g., corridor travel time), but also the local traffic phenomena (e.g., bottlenecks, queues, speed-flow patterns) should be reproduced in the simulation to ensure a good representation of the real-world traffic conditions.}
Recently studies revealed the shifted attention to evaluating the interaction between CACC and non-equipped vehicles, as well as the operational strategy that can harness the benefits of CACC in mixed traffic conditions. 
The coordination among CAVs to form vehicular platoons as an operational strategy has shown potential in boosting CACC benefits when the MPR is low. However, its potential negative impact due to induced lane change has not been investigated at the vehicle trajectory level.

\section{Evaluation Method}
\label{sect:framework}

\subsection{Simulation of Human Driving Behavior}
Human drivers can take into account multiple stimuli (e.g., brake lights, next-nearest neighbors, etc.) with anticipation of the situation for the next few seconds \citep{treiber2013traffic}. All of these aspects can be formulated into psycho-physiological models, such as the Gipps' model \citep{Gipps1981A} and the Wiedemann model \citep{Wiedemann1974, wiedemann1991modelling, ptv2018ptv}. 
\begin{figure} [h]
	\centering
	\includegraphics[width=\linewidth]{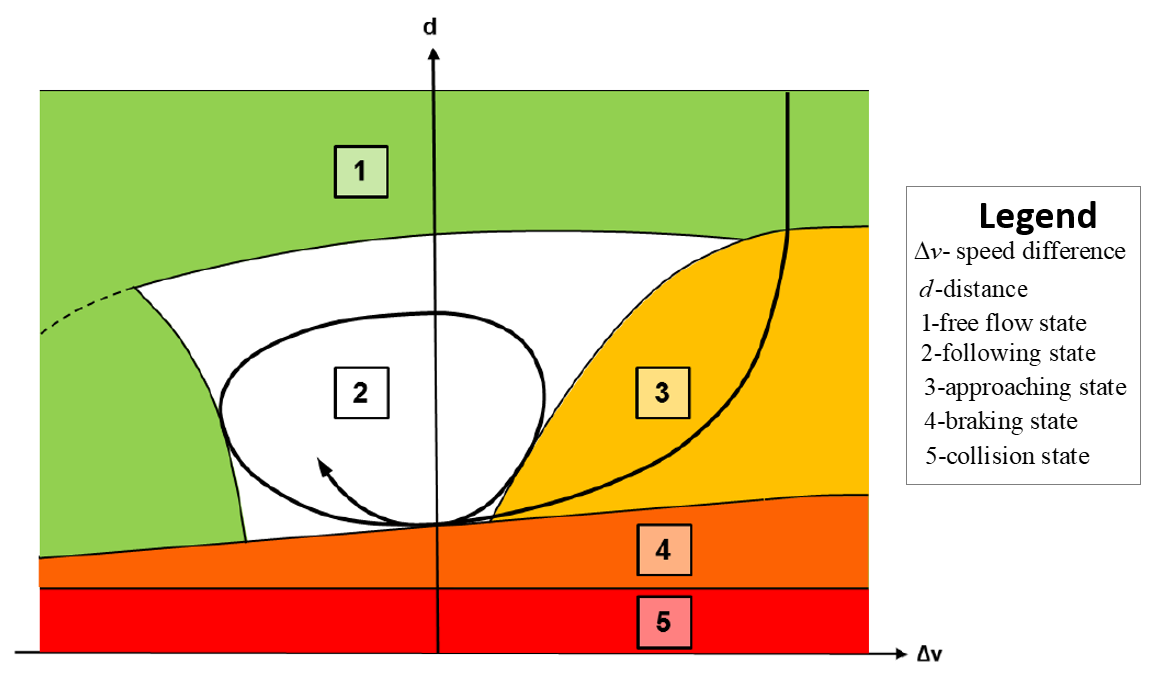}   
	\caption{\textcolor{r3}{Wiedemann 1974 model and car-following state \citep{ptv2018ptv}}}
	\label{fig:w99Model}
\end{figure}

The Wiedemann model is used by Vissim as the default car-following model. It includes the tactical driving behavior, which carries certain planning in advance with a temporal horizon (multiple time steps) and a spatial horizon that is beyond the neighbor vehicles \citep{fellendorf2010microscopic}. 
\textcolor{r3}{Such spatial horizon is an augmentation of the initial Wiedemann 74 model in the Vissim car-following behavior, which include ranges of the look-ahead and look-back distances and the numbers of interaction objects/vehicles \citep{ptv2018ptv}}.
As illustrated in Fig. \ref{fig:w99Model}, there are five different driving states in the Wiedemann model:
\begin{enumerate*}[label=\roman*)]
\item free driving,
\item approaching,
\item following, 
\item braking, and
\item collision.
\end{enumerate*} 
\textcolor{r3}{The acceleration is primarily determined by the speed difference (x-axis) and headway/distance (y-axis) to the preceding vehicle for each of the five driving states.} The Wiedemann-99 model, suitable for freeway application, has ten calibratable parameters that can constitute a wide range of driver behaviors. Therefore, the Wiedemann model has to be calibrated according to specific traffic stream data \citep{higgs2011analysis}, as it was initially developed on limited data \citep{reiter1994empirical}. The objective of the calibration process is to minimize the difference between the measured driving behavior in the field and the simulated driving behaviors.

When it comes to lateral control, necessary lane change and free lane change are the two types of lane changes in the \textcolor{r3}{lateral movement model of Vissim}. The former focuses on the hard constraint of the lane change (e.g., lane drop). The latter type is the focus of this study. Such lane change is performed when more space or higher speed is desired for a vehicle. As such, the safety distance plays an important role in lane change behavior, and it is determined based on: 
\begin{enumerate*}[label=\roman*)]
\item the speed of the vehicle changing lane, and 
\item the approaching speed of the following vehicle \citep{ptv2018ptv}
\end{enumerate*} 

\subsection{Quantifying Impact to Human Driver}
\begin{figure*} [h]
	\centering
	\includegraphics[width=\textwidth]{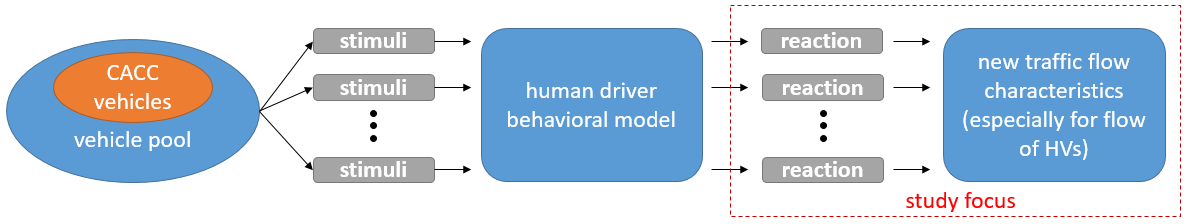}   
	\caption{Potential impact on non-equipped/human-driven vehicles} 
	\label{fig: impactEvaluationMethod}
\end{figure*}

A before-and-after comparison is the most straightforward way to assess the changes that are brought by CAVs.  Fig. \ref{fig: impactEvaluationMethod} illustrates the study methodology. First, human driving behavior is calibrated by multiple sources of data that were collected from the roadway segment of interest. \textcolor{r3}{Not only the global traffic characteristics, such as corridor travel time, but also the local traffic patterns (e.g., bottlenecks, queues) were replicated in the simulation.} We then treat the car-following model as a plant model. The input is localized traffic conditions, and the output is the reactions of HVs. On a collective level, the traffic flow characteristics and vehicle trajectories are analyzed.

\subsection{Simulation of CAV Driving Behavior}
Longitudinal control (or car following) and lateral control are the two main components for simulating the behavior of CACC. 
\textcolor{re}{The intelligent driver model (IDM) and its variants have been used to design ACC/CACC controller that resembles human-like car-following behaviors \citep{Wang2019Benefits, Kesting2008adaptive, Talebpour2016Modeling, Spiliopoulou2017Exploitation, gueriau2016assess}.}
The Enhanced Intelligent Driver Model (E-IDM) \citep{Kesting2010} is adopted as the longitudinal control model, which is expressed in Eq. (\ref{eq: idm}).  

\begin{subequations}\label{eq: idm}
\begin{equation}
\ddot{x} =\begin{cases}
a[1-(\frac{\dot{x}}{\dot{x_{des}}})^{\delta }- (\frac{s^{*}(\dot{x}, \dot{x}_{lead})}{s_{0}})] & \text{ if } x=  \ddot{x}_{IDM} \geq \ddot{x}_{CAH} \\ 
 (1-c)\ddot{x}_{IDM} + c[\ddot{x}_{CAH} + b \cdot tanh ( \frac{\ddot{x}_{IDM} - \ddot{x}_{CAH}}{b})] & \text{otherwise} 
\end{cases}\\
\end{equation}
\begin{equation}
s^{*}(\dot{x}, \dot{x_{lead}}) = s_{0} + \dot{x}T + \frac{\dot{x}(\dot{x} - \dot{x_{lead}})}{2\sqrt{ab}} \\
\label{eq: minDistCal}
\end{equation}
\begin{equation}
\ddot{x}_{CAH} =
\begin{cases}
\frac{\dot{x}^{2} \cdot \min(\ddot{x}_{lead}, \ddot{x})}{\dot{x}_{lead}^{2}-2x \cdot \min(\ddot{x}_{lead}, \ddot{x})} & 
\dot{x}_{lead} (\dot{x} - \dot{x}_{lead}) \leq -2x \min(\ddot{x}_{lead}, \ddot{x})  \\
\min(\ddot{x}_{lead}, \ddot{x}) - \frac{(\dot{x}-\dot{x}_{lead})^{2} \Theta (\dot{x}- \dot{x}_{lead})}{2x}  & \text {otherwise}
\end{cases} 
\label{eq: CAH}
\end{equation}
\end{subequations}
where, $a$ is the maximum acceleration; $b$ is the desired deceleration; $c$ is the coolness factor; $\delta$ is the free acceleration exponent; $\dot{x}$ is the current speed of the subject vehicle;  $\dot{x}_{des}$ is the desired speed,  $\dot{x}_{lead}$ is the speed of the lead vehicle; $s_{0}$ is the minimal distance; $\ddot{x}$ is the acceleration of the subject vehicle; $\ddot{x}_{lead}$ is the acceleration of the lead vehicle; $\ddot{x}_{IDM}$ is the acceleration calculated by the original IDM model \citep{Treiber2000}. The minimal distant can be calculated by Eq. (\ref{eq: minDistCal}), where  $T$ is the desired time gap; and $\ddot{x}_{CAH}$ is the acceleration calculated by the CAH component as shown in Eq. (\ref{eq: CAH}),{eq: CAH} where $\Theta$ is the Heaviside step function. The parameters used for the E-IDM is shown in Table \ref{table:simParameter}. The CAV uses 0.6 s as the desired time gap when following another CAV and 0.9 s otherwise. The desired speed is uniformly distributed between 96 and 105 $km/h$, and it varies by individual CAV.

\begin{table}[!ht]
\centering
\caption{EIDM Parameters} 
\begin{tabular}{ccccccccccc}
\hline \hline
Parameter & $T$ & $s_{0}$ & $a$ & $b$ & $c$ & $\theta$ & $\dot{x}_{des}$ & $\phi_{max}$ & $D_f$ & $D_r$ \\ \hline
value & 0.6 or 0.9 s  & 1$m$ & 2$m/s^{2}$ & 2$m/s^{2}$ & 0.99 & 4 & 96$\sim$105 $km/h$ & 5 & 20 $m$ & 20 $m$   \\
\hline
\end{tabular}
\label{table:simParameter}
\end{table}

We assume all the CAVs are equipped with automated longitudinal control. Each CACC vehicle is able to detect the surrounding traffic and discern CAVs from HVs. Three cases, as shown in Table \ref{table:scenario}, are tested.

\begin{table}[H]
\center
\caption{Simulation cases}
\begin{tabular}{l|ll}
\hline \hline
\textbf{Case} & \textbf{Longitudinal Control} & \textbf{Lateral Control}  \\ \hline
Base (no CAVs) & calibrated Wiedemann & \textcolor{re}{Vissim} \\
Ad hoc coordination & E-IDM & \textcolor{re}{Vissim}\\
Local coordination & E-IDM &  gap acceptance-based \citep{lee2014mobility}\\
\hline
\end{tabular}
\label{table:scenario}
\end{table}

\begin{figure}[h]
\begin{minipage}[h]{\linewidth}
\centering
\subfloat[a free-agent CACC vehicle lane change to join an existing platoon]{\includegraphics[scale=0.5]{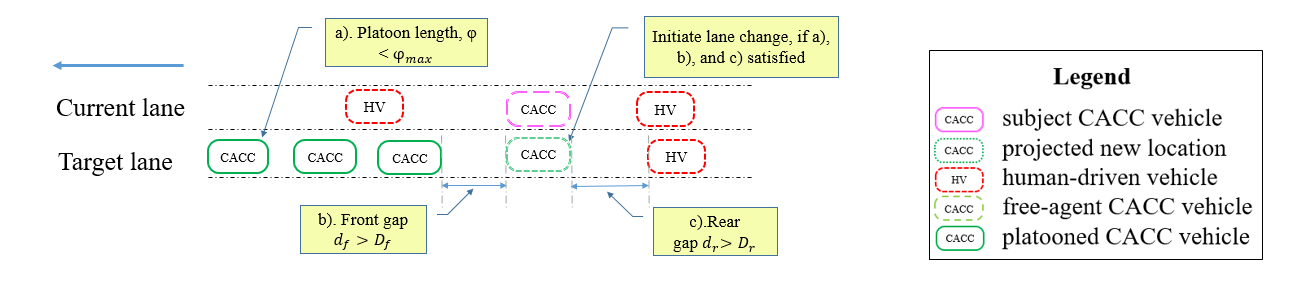}}
\end{minipage} \par
\begin{minipage}[h]{\linewidth}
\centering
\subfloat[selected CACC vehicle to create a gap for an incoming free-agent CACC vehicle]{\includegraphics[scale=0.5]{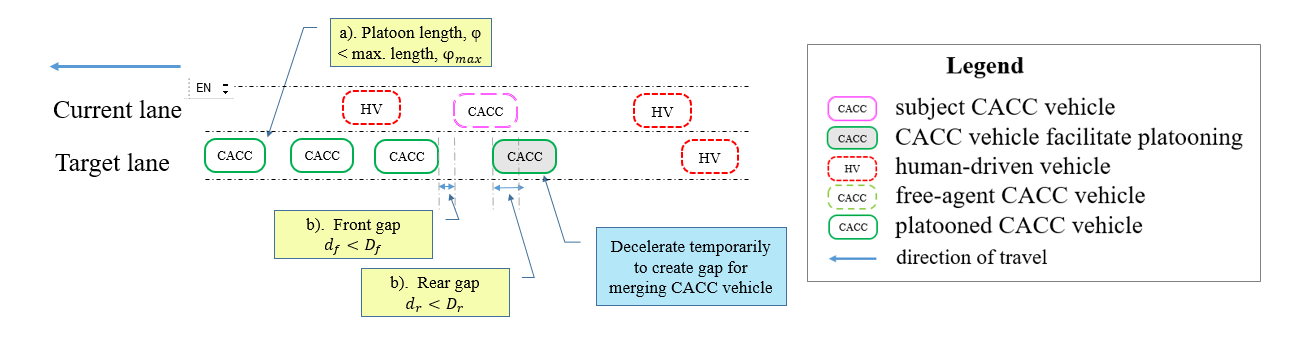}}
\end{minipage}
\caption{\textcolor{re}{Local coordination strategy}}
\label{fig:localCoordStrat}
\end{figure}

\textcolor{re}{
The lateral control for CACC vehicles in platoon formation was developed by \cite{lee2014mobility}, which has been adopted in multiple studies \citep{Zhong2017a, NAP25366,OSDAP2015}. As demonstrated in Fig. \ref{fig:localCoordStrat}, a free-agent CACC vehicle actively communicates with its surrounding vehicles/platoons for platoon formation. Upon the discovery of a platoon opportunity, the free-agent CACC vehicle used its on-board sensors to determine the front and rear gaps based on its lateral projection to the target lane. The lane change is only initiated if both front and rear gaps are above the predetermined thresholds. When the incoming vehicle needs to joint the platoon in the middle, the nearest member within the platoon will decelerate temporarily to create sufficient gaps to facilitate the lane change of the incoming CACC vehicle. The flow chart of the algorithm is provided in Fig. \ref{fig:leeLcAlgo} in \ref{sect:flowchart}.
}

\subsection{Case Study Network}
\begin{figure} [H]
	\centering
	\includegraphics[width=0.8\textwidth]{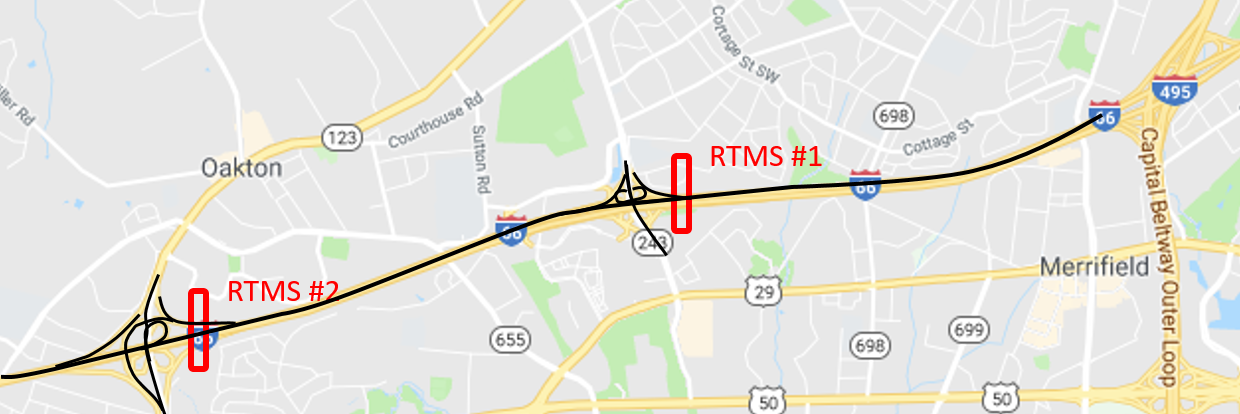}   
	\caption{I-66 simulation testbed} 
	\label{fig: aaControlConfig}
\end{figure}

In this study, an 8-km (5-mi) segment (Fig. \ref{fig: aaControlConfig}) of Interstate Highway I-66 located outside of the beltway (I-495) of Washington D.C. is considered. This freeway segment has recurring congestion during weekdays, specifically in the eastbound direction in the morning and the westbound direction in the afternoon. The roadway has four lanes in each direction. The leftmost lane is an HOV 2+ lane with a peak volume of 1,500 vph per lane \citep{lu2014freeway}. Currently, there is no physical barrier between the HOV lane and the adjacent general purpose (GP) lane. The calibration was conducted with three independent data sources:  INRIX travel time, remote traffic microwave sensor data (RTMS), and video camera data to ensure a realistic representation of the real-world traffic conditions \citep{STOLT4, Li2019High}. The first two were the primary sources of fine-tuning the simulation network.
\textcolor{r2}{ The calibration for the Vissim network is further elaborated in \ref{sect:appCalibration}}.

In anticipation of the increasing traffic demand over time, we assume a 30\% increase in the traffic demand from the baseline of the calibrated network. \textcolor{r3}{The Wiedemann model in Vissim has stochastic terms, which is implemented via simulation random seed, provided all other parameters unchanged. Therefore, five replications was simulated for each deployment scenario with different random seeds to account for the stochasticity of the traffic conditions. Across the scenarios, the same set of random seeds is used to ensure fair comparison.} The duration of the simulation is 3,900 s, the first 300 s of which is used to load the network with traffic. No data is collected during the first 300-s period.
The evaluation of the impact is based on the following assumptions:
\begin{itemize}
\item A low-level vehicle controller for longitudinal and lateral control is available. 
\item A calibrated driving behavior model in Vissim with real-world data constitutes a good representation of a subset of the human driver population.
\item Vehicle-to-vehicle (V2V) communication is perfect (no interference or packet loss).
\item \textcolor{re}{ Human drivers do not differentiate CAVs and another HVs in their surrounding,  and the behavioral adaptation for CAV behavior \citep{gouy2013behavioural} is absent}. 
\item 1.3 times of the original demand is adopted in anticipation of lead-time for CAV deployment.
\end{itemize}

\section{Results \& Discussions}
\label{sect:result}
The vehicle trajectory data are collected in every 0.5 s. Each simulation replication contains approximately seven million lines of vehicle records, which include not only the vehicle dynamic data but also the interaction states (e.g., driving state and interacting vehicles). A total number of 45 replications are conducted. 

\subsection{Network Performance}
\begin{figure}[h]
\begin{minipage}[h]{.5\linewidth}
\centering
\subfloat[productivity]{\includegraphics[scale=0.5]{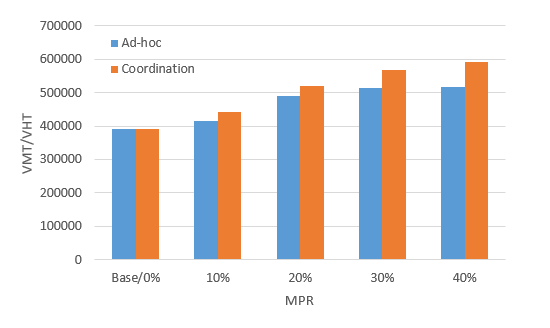}}
\end{minipage} 
\begin{minipage}[h]{.5\linewidth}
\centering
\subfloat[throughput]{\includegraphics[scale=0.5]{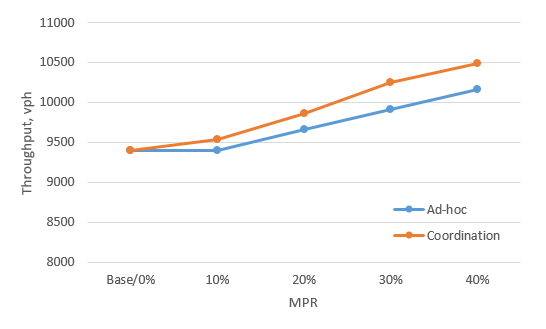}}
\end{minipage}\par
\caption{Network performance}
\label{fig:netPems}
\end{figure}

Fig \ref{fig:netPems}(a) shows the ratio of vehicle miles traveled (VMT) and vehicle-hours traveled (VHT). VMT is the output of a transportation system, whereas VHT is considered the input to a transportation system. The ratio of VMT and VHT ratio is referred to as Q \citep{Caltrain2013aGuide}, which represents the output of a transportation system with the unit value of the input. In short, the higher the value of the Q, the more productive a transportation system is. Either of the ad hoc coordination or local coordination exhibits an increasing trend as the MPR increases. It is notable that the benefits gained by the ad hoc coordination show a diminishing increase after 30\% MPR.  In comparison, local coordination displays a linear increasing pattern for the performance gain. 

When it comes to network throughput,  Fig. \ref{fig:netPems}(b) shows that the ad hoc coordination does not increase the network throughput at 10\% MPR-the throughput remains as 9,398 vph. Above 10\% MPR, the throughput for ad hoc coordination increases with a linear pattern, which matches the underlying operational implication of ad hoc coordination. The throughput reaches the highest 10,167 vph at 40\%. With local coordination, additional throughput is observed even at 10\% MPR. In addition, the slope of the throughput curve for local coordination is greater at the MPR range between 10\% and 30\% than other tested MPR values.  It also shows that the rate of throughput increment starts to decrease beyond 30\% MPR.  Moreover, the vertical distance indicates the magnitude that the local coordination outperforms the ad hoc coordination at each level of MPR. The greatest difference is observed at 30\% MPR.

\subsection{Hard Braking Observations}
\begin{figure}[h]
\begin{minipage}[h]{.5\linewidth}
\centering
\subfloat[hard braking observation CDF]{\includegraphics[scale=0.23]{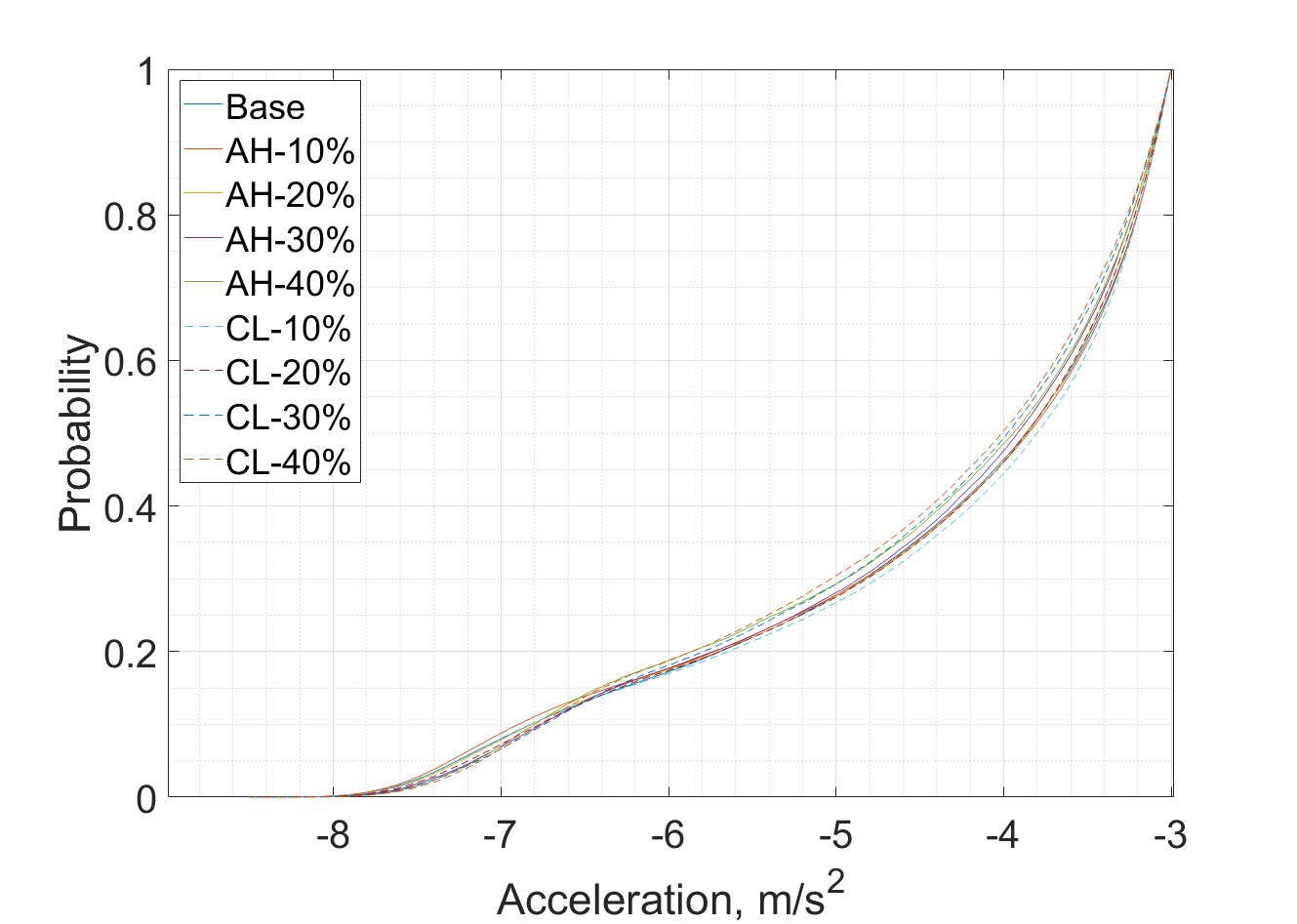}}
\end{minipage}
\begin{minipage}[h]{.5\linewidth}
\centering
\subfloat[hard braking observation]{\includegraphics[scale=0.23]{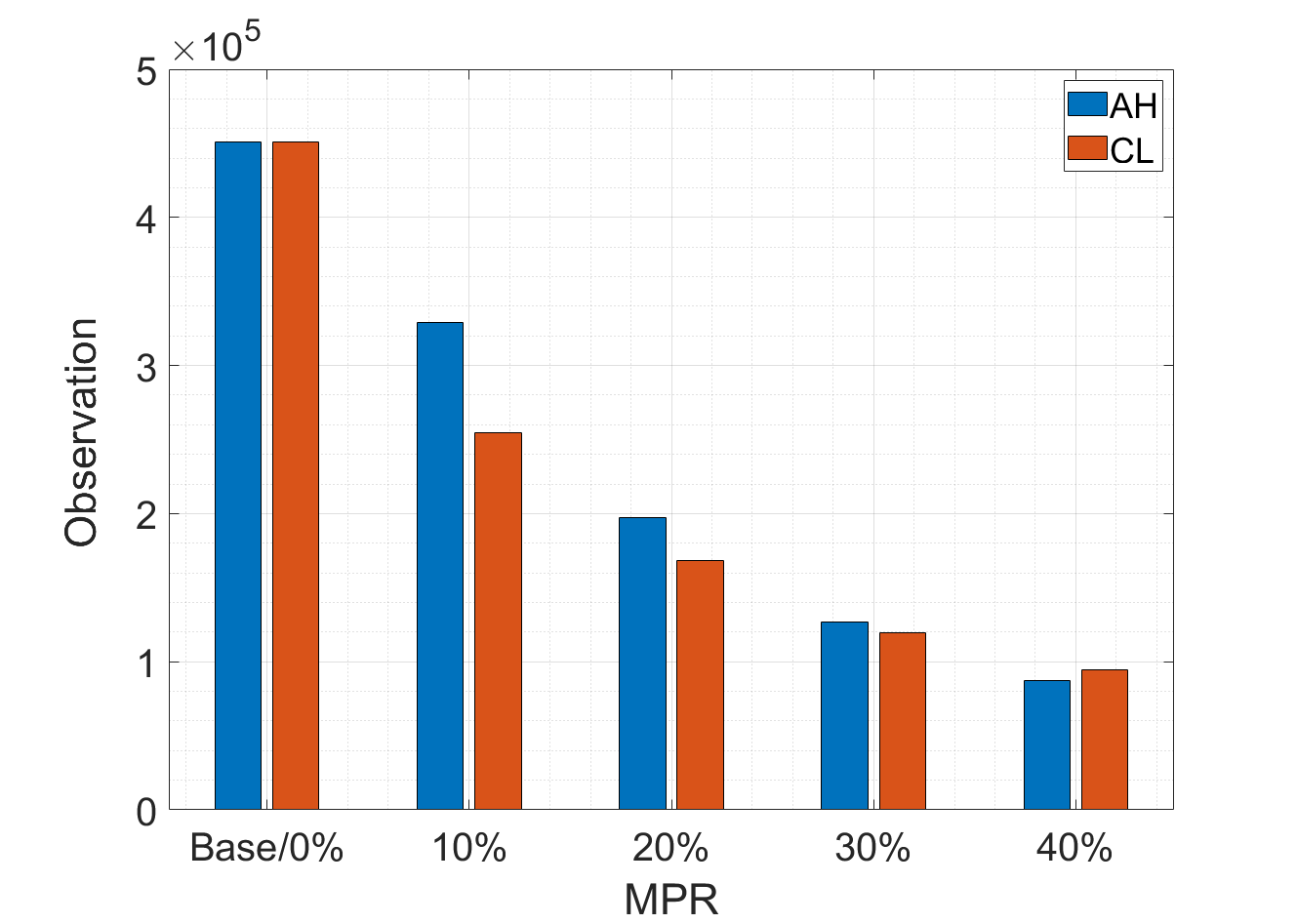}}
\end{minipage}
\caption{Hard braking observation when interacting with HVs}
\label{fig:hdCdfHv}
\end{figure}

Abrupt braking is an indication of a hazardous traffic situation that drivers respond to \citep{bagdadi2011jerky}. Hard braking observations are recorded when the acceleration of a vehicle is less than -3 $m/s^2$.  Here, our focus is on the HVs. Based on the interacting vehicle, there are two types of hard braking. The first type occurs when an HV interacts with another HV (HV-HV type interaction); whereas the second type occurs when an HV interacts with a CAV (HV-CAV type interaction). Fig. \ref{fig:hdCdfHv}(a) shows the hard braking observations for HVs when they interact with other HVs.  Similar patterns of the cumulative distribution functions (CDFs) for hard braking are observed across the testing scenarios.  Fig. \ref{fig:hdCdfHv}(b) shows the number of observations  for each scenarios. The primary factor for the decreasing trend is the reduction of HVs in the traffic stream. The linear trend also infers that the likelihood of hard braking remains at the same level. 

\begin{figure}[h]
\begin{minipage}[h]{.5\linewidth}
\centering
\subfloat[hard braking event CDF]{\includegraphics[scale=0.24]{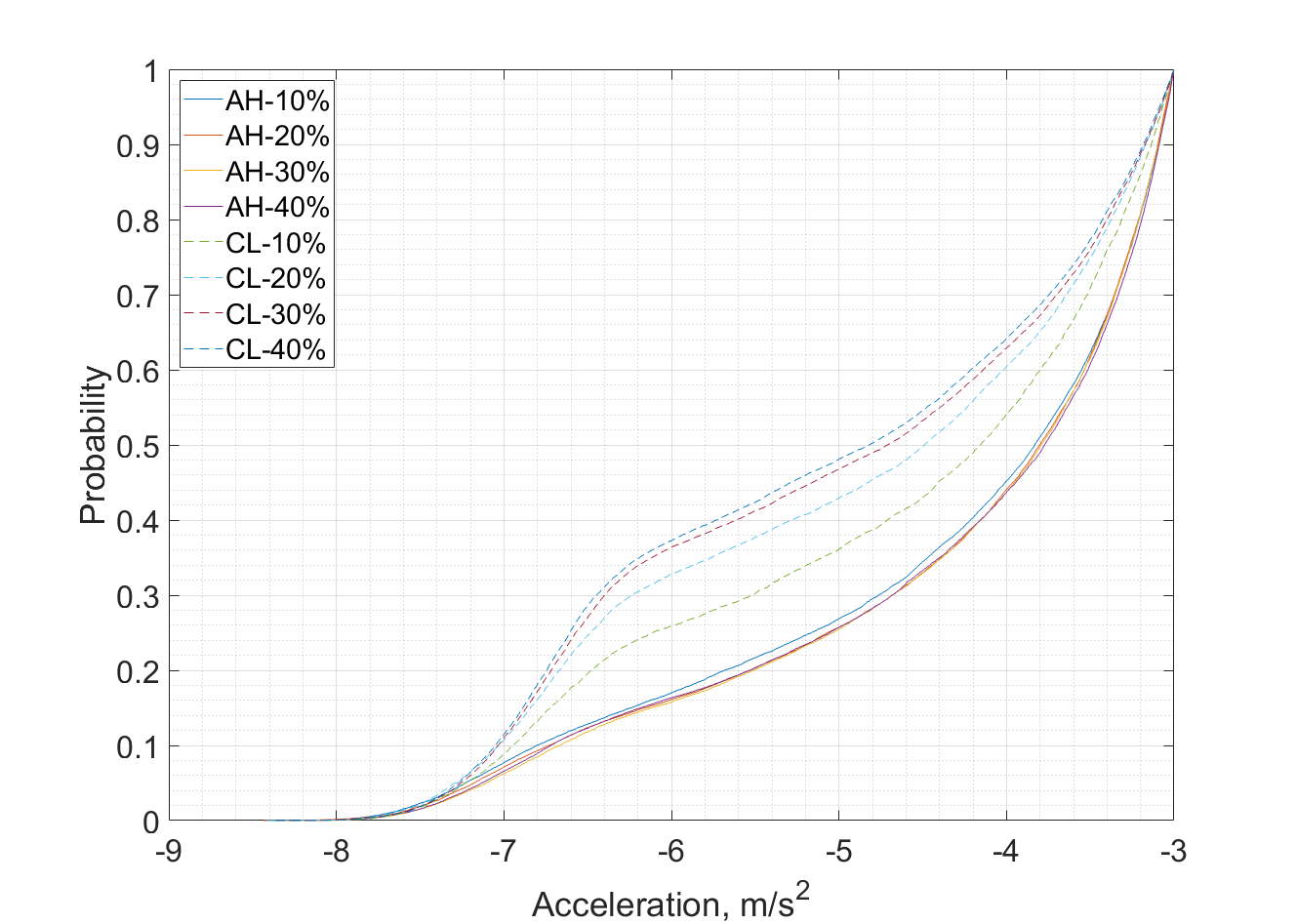}}
\end{minipage}
\begin{minipage}[h]{.5\linewidth}
\centering
\subfloat[hard braking event observation]{\includegraphics[scale=0.24]{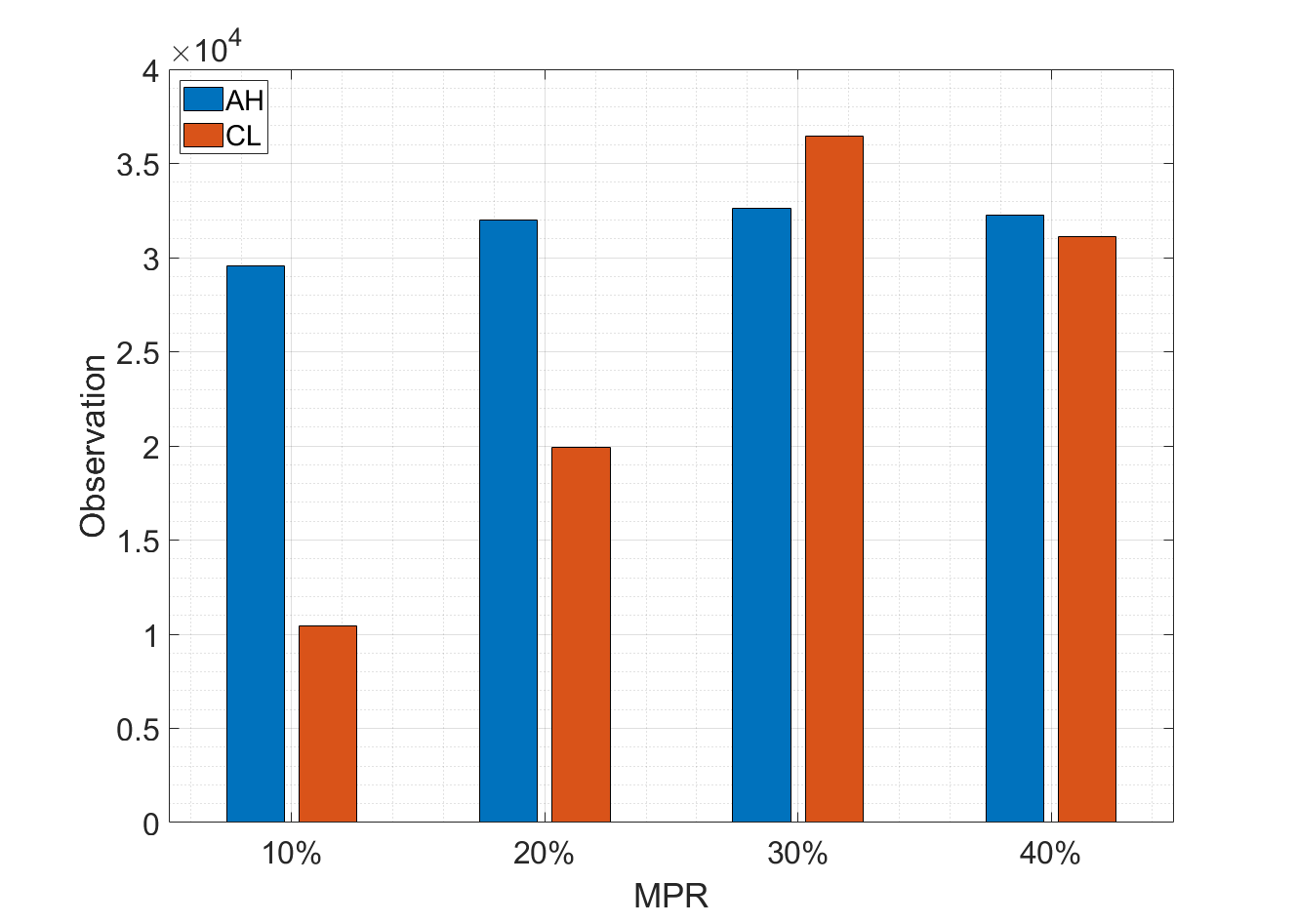}}
\end{minipage}\par
\caption{Hard braking event when interacting with CAVs}
\label{fig:hdCdf}
\end{figure}

Fig. \ref{fig:hdCdf}(a) shows the CDFs of the hard breaking observations recorded for HVs when they interact with CAVs. The CDFs show two distinctive patterns between two coordination strategies. In the ad hoc coordination case, the CDFs are with similar distributions. On the other hand, the CDFs of local coordination are more sensitive to MPR. The probability of hard braking between -6.5 $m/s^2$ and -3.5 $m/s^2$ increases drastically even at 10\% MPR. The occurrence of hard braking events keeps at the same level in ad hoc coordination; whereas the occurrence of coordination strategy shows an increasing trend until reaching 30\% MPR. 
The number of observations is shown in Fig. \ref{fig:hdCdf}(b). Both strategies exhibit an increasing trend, then a declining trend after 30\% MPR. With the same amount of CAVs, the hard braking is more sensitive to MPR in local coordination than that in the ad hoc coordination.

Two-sample Kolmogorov-Smirnov (K-S) test is adopted to further analyze the CDFs. The K-S test is a powerful tool for testing the hypothesis of whether two random samples have been drawn from the same population \citep{goodman1954kolmogorov}. It is a non-parametric test where no assumption is made regarding the distribution of the variables \citep{young1977proof}. The null hypothesis ($H_0$) of the K-S test is that the comparing two sample sets are from the same continuous distribution. Table \ref{table: kmTest_a} shows the result of the two-sample K-S test for the hard braking CDFs for HVs. The value of ``1" represents that the null hypothesis is rejected at a 5\% significance level. As shown, any pair of the scenarios rejects the null hypothesis and accept the alternative hypothesis that the two samples are not from the same distribution. The observations only fail to reject the $H_0$ in the ad hoc coordination case at 20\% and 30\% MPR .

\begin{table}[h]
\centering
\caption{Two-sample K-S test for Hard Braking} 
\begin{tabular}{c|cccccccc}
\hline \hline
 & A-10\% & A-20\% & A-30\% & A-40\% & C-10\% & C-20\% & C-30\% & C-40\% \\ \hline
A-10\% & - & 1 & 1 & 1 & 1 & 1 & 1 & 1 \\
A-20\% & 1 & - &  \textbf{0} & 1 & 1 & 1 & 1 & 1 \\
A-30\% & 1 & \textbf{0} & - & 1 & 1 & 1 & 1 & 1 \\
A-40\% & 1 & 1 & 1 & - & 1 & 1 & 1 & 1 \\ \hline
C-10\% & 1 & 1 & 1 & 1 & - & 1 & 1 & 1 \\
C-20\% & 1 & 1 & 1 & 1 & 1 & - & 1 & 1 \\
C-30\% & 1 & 1 & 1 & 1 & 1 & 1 & - & 1 \\
C-40\% & 1 & 1 & 1 & 1 & 1 & 1 & 1 & -\\
\hline
\end{tabular}
\label{table: kmTest_a}
\end{table}

\subsection{Safety Surrogate Assessment Measure}
\begin{figure}[H]
\begin{minipage}[H]{0.5\linewidth}
\centering
\subfloat[HV-HV]{\includegraphics[scale=0.46]{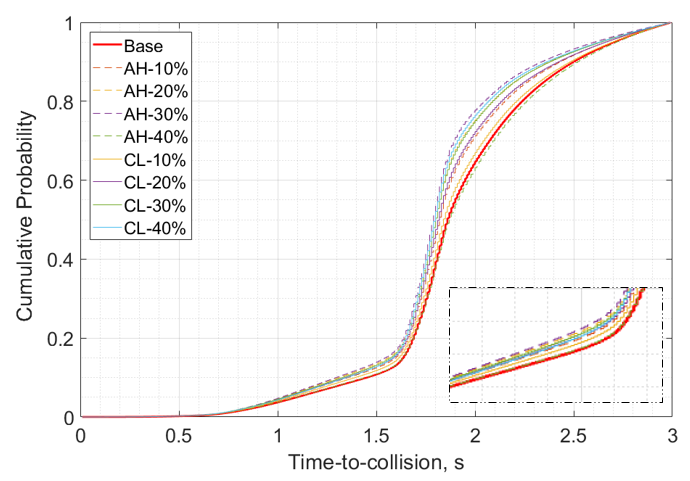}}
\end{minipage}  
\begin{minipage}[H]{0.5\linewidth}
\centering
\subfloat[HV-CAV]{\includegraphics[scale=0.46]{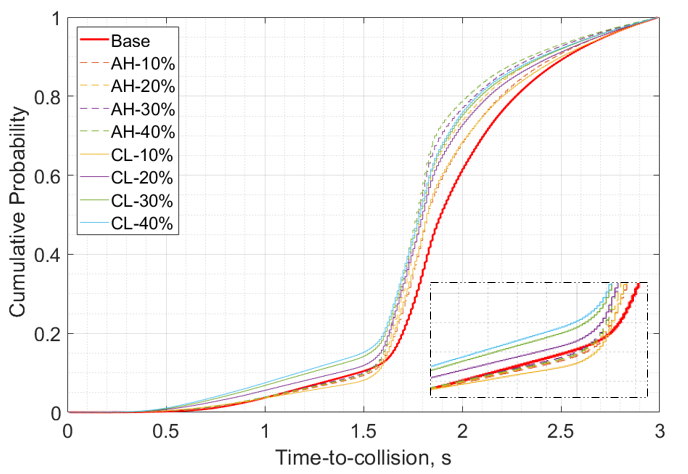}}
\end{minipage}
\caption{\textcolor{re}{Time-to-collision (TTC) CDF}}
\label{fig:ttc}
\end{figure}

\textcolor{re}{
The Surrogate Safety Assessment Model (SSAM) is a tool for identifying traffic conflicts from vehicle trajectories \citep{gettman2008surrogate}.  
The time to collision (TTC) less than 3 s to a preceding vehicle for each HV is extracted from the trajectory data. Fig. \ref{fig:ttc} (a) shows the TTC for HV-HV type interaction and  Fig. \ref{fig:ttc} (b) shows the HV-CAV interaction. Like hard braking, we performed the two-sample K-S test for each pair of CDFs and found that all the comparisons reject the null hypothesis. Then we examined the CDF portion where TTC is under 1.5 s. For the HV-HV type interaction, the cumulative probability increase along with the MPR. For the HV-CAV type interaction, the probability of TTC less than 1-s is higher with local coordination except at 10\% MPR, and the cumulative probability is higher as the MPR grows. Under ad hoc strategy, the cumulative probability slightly increases when TTC is less than 0.8 s, and then they fall below the Base CDF curve between 1 s and 1.6 s. 
}

\subsection{Lane Change Activities}
\begin{figure} [h]
	\centering
	\includegraphics[width=.8\linewidth]{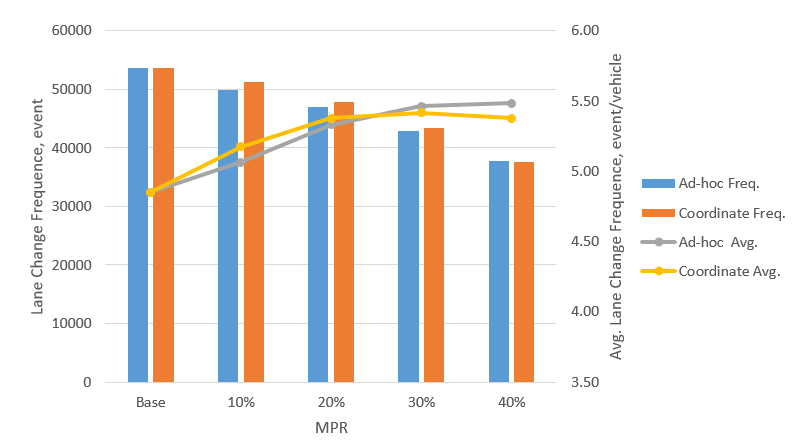}   
	\caption{Lane change activity of HVs} 
	\label{fig: lcFreqHV}
\end{figure}

Fig. \ref{fig: lcFreqHV} shows the accumulative lane change recorded at every 0.5 s for all HVs. 
\textcolor{re}{
As mentioned previously, the Vissim has a necessary lane change and free lane change in its default model. 
}
The lane change activity decreases as the MPR of CACC increases in either of the coordination strategies. Local coordination is marginally higher than the ad hoc one when the MPR is low. At 40\% MPR, they reach the same level of lane change activity. However, recall that the number of HV within the network decreases as the MPR of CACC increases. The average lane change frequency for each HV is plotted in Fig.\ref{fig: lcFreqHV} as well. The local coordination strategy shows a higher average lane change frequency at 10\% and 20\% MPR. The average lane change frequency peaks at 30\%, then reduced to 5.38 from 5.42 per vehicle. On the contrary, the increasing trend for ad hoc coordination keeps increasing and reach 5.46 and 5.48 per vehicle at 30\% and 40\%, respectively. 30\% is the stationary point when it comes to average lane change frequency.
\textcolor{re}{The surrounding CAV could induce HVs for more lane change for greater space or higher speed. Some examples for the reaction of HVs include: 
\begin{enumerate*}[label=\roman*)]
\item the high traffic density resulted from the closely-coupled CACC platoon, 
\item the short following distance under which a CAV can be safely operated,
\item the narrower gap for CAV lane change could encroach the desired headway for HV, hence prompts HV's lane change
\end{enumerate*}
}

\subsection{Interaction State}
The Wiedenmann model assumes five basic interaction states for a human driver. The Braking and Collision states are often coded as ``Brake BX" and ``Brake AX," respectively. The former represents breaking performed before reaching the safety distance, whereas the latter represents braking performed after reaching the safety distance. The definition of each interaction state is listed in Table \ref{table:intactState}.
 The transition to one state to another is triggered by a threshold, which can be expressed as a function of the speed difference and distance \citep{ptv2018ptv}.

\begin{table}[H]
\centering
\caption{Wiedenmann's car-following interaction state} 
\resizebox{0.9\textwidth}{!}
{
\begin{tabular}{p{2in}|p{5in}} 
\hline \hline 
\textbf{Parameter} & \textbf{Definition}\\ \hline 
Free &  a vehicle is not affected by any relevant preceding vehicle. It tries to drive at its desired speed\\ \hline 
Close-up & a vehicle is reaching an obstacle (e.g., signal head, vehicle, conflict area)  \\ \hline
Follow & a vehicle tries to follow a leading vehicle\\ \hline
Brake BX & braking at the desired safety distance before reaching the safety distance\\ \hline
Brake AX & braking after reaching the safety distance\\ 
\hline   
\end{tabular}
}
\label{table:intactState}
\end{table}

\begin{figure}[H]
\begin{minipage}[h]{.5\linewidth}
\centering
\subfloat[ad hoc coorindation]{\includegraphics[scale=0.5]{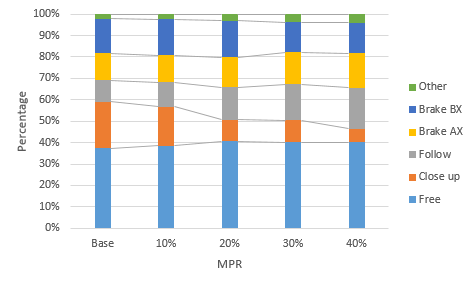}}
\end{minipage}
\begin{minipage}[h]{.5\linewidth}
\centering
\subfloat[local coordination]{\includegraphics[scale=0.5]{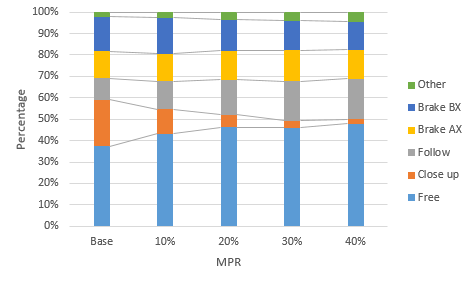}}
\end{minipage}
\caption{Composition of interaction state for HVs}
\label{fig:driStateComp}
\end{figure}

The high-resolution trajectory data contains the interaction state information, which provides insights for the influence of CAVs on HVs. By analyzing the interaction data, we can quantify the possible influence of CAV. The composition of the interaction state is shown in Fig. \ref{fig:driStateComp}.  The non-major interaction states are grouped into a new category-``Other". The percentage of ``Free" state for HVs is higher in the local coordination strategy at each level of MPR compared to its counterparts. In the ``Base" case, on average 37\% of the time the HVs are driving under ``Free" condition. The introduction of CAVs with ad hoc strategy only marginally increases the percentage of ``Free" condition for HVs. When implementing the Coordination strategy, the percentage of ``Free" condition gains a greater increase, and it reaches 48\% at 40\% MPR. 

\begin{figure}[H]
\begin{minipage}[h]{.5\linewidth}
\centering
\subfloat[non-braking following states]{\includegraphics[scale=0.65]{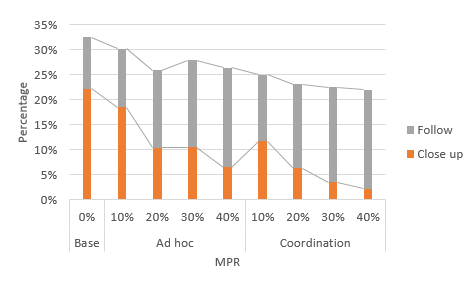}}
\end{minipage} 
\begin{minipage}[h]{.5\linewidth}
\centering
\subfloat[braking states]{\includegraphics[scale=0.65]{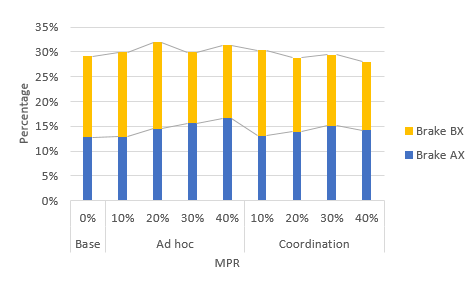}}
\end{minipage}
\caption{Interaction state for HV}
\label{fig:interactionStateForHV}
\end{figure}

The ``Close-up" and ``Follow" states are isolated and shown in Fig. \ref{fig:interactionStateForHV}(a). The percentage of the ``Close-up" state exhibits a decreasing trend as the MPR increases. With the local coordination strategy, the time HV spent in the ``Close-up" state decline more rapidly as the MPR increases. It is also observed that HVs spend more time in the ``Follow" state when CAVs use local coordination strategy.  The increased percentage of the ``Follow" state indicates a shorter time headway on average. Recall in Fig. \ref{fig:netPems}(b) that the throughput is increased with CAVs. Therefore the density of the traffic is increased, resulting in a shorter headway on average and more time in Follow state.
Fig. \ref{fig:interactionStateForHV}(b) details the compositions of the two braking states. The ``Brake AX" state has more significant safety implications, as it reflects the braking that is performed after a reaching safety distance. \textcolor{re}{The observations for the ``Brake AX" reach the highest 17\% (at 40\% MPR) and 15\% (at 30\% MPR) for ad hoc coordination and local coordination strategies, respectively. The ''Brake AX" increases monotonically in Ad hoc case; whereas it decreases back to 14\% after the peak.}

\begin{figure}[h]
\begin{minipage}[h]{\linewidth}
\centering
\subfloat[speed difference-headway plot]{\includegraphics[scale=0.28]{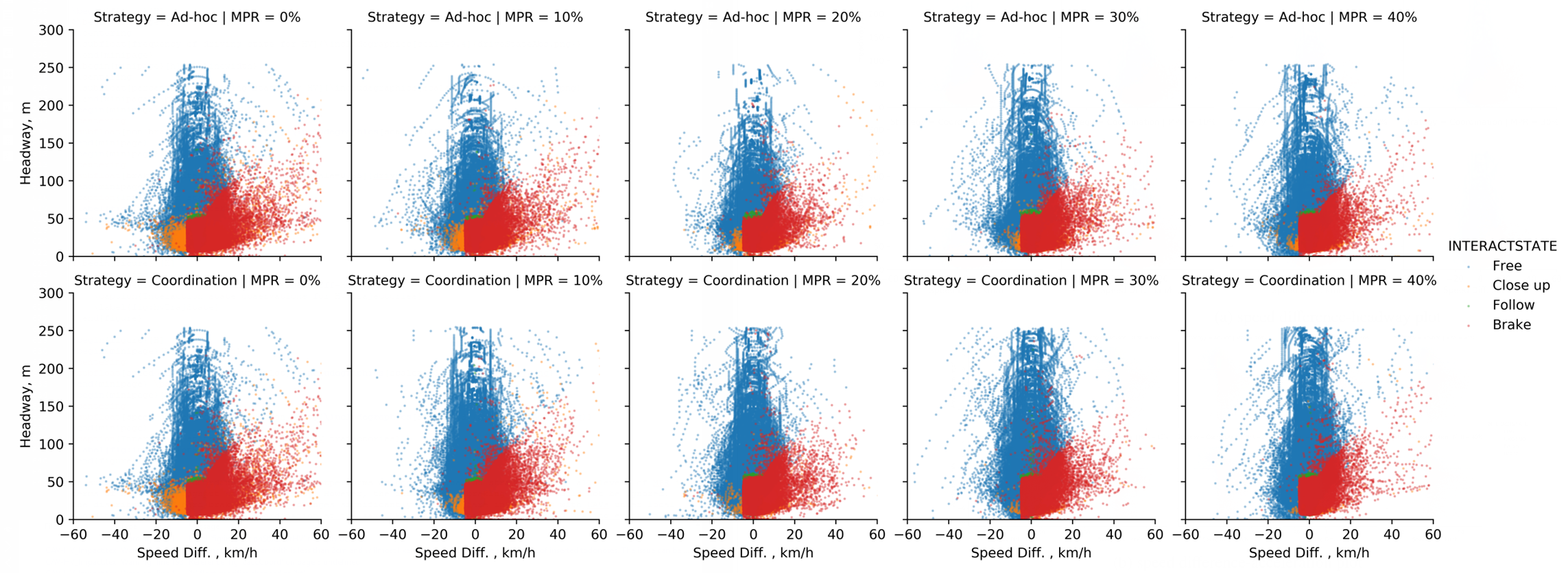}}
\end{minipage} \par
\begin{minipage}[h]{\linewidth}
\centering
\subfloat[speed difference-acceleration plot]{\includegraphics[scale=0.28]{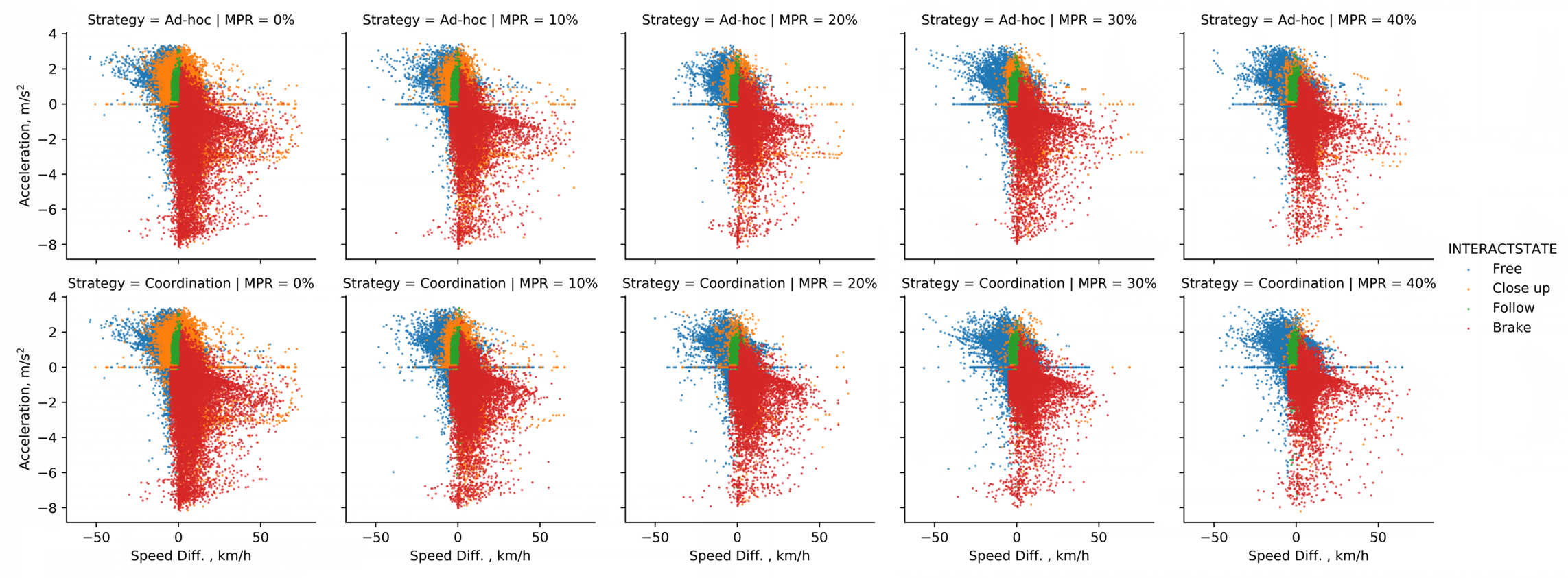}}
\end{minipage}
\caption{Interaction state observations for HV trajectories}
\label{fig:hvTrajState}
\end{figure}

Fig. \ref{fig:hvTrajState}(a) shows the speed difference-headway plot, which imitates the theoretical plot in Fig. \ref{fig:w99Model}. Each vehicle slightly varies in the separation of the five following states, depending on the random seed. Due to a large number of observations, overlapping among sample points is hard to avoid, especially for the same following state sharing the same color. Fig. \ref{fig:hvTrajState}(b) displays the speed difference-acceleration plot. It shows that the disappearance of the observation of the  ``Close-up" state as MPR increases for both strategies. The reduction is greater in the local coordination case, which is consistent with the trend shown in Fig. \ref{fig:interactionStateForHV}(a).

In summary, the local coordination strategy outperforms ad hoc coordination strategy across all levels of MPR in terms of network throughput and productivity. However, the induced hard braking for HV should not be overlooked. The distribution of hard braking for HVs changes significantly when they are interacting with CAVs under local coordination. Compared to ad hoc coordination at the same MPR, the probability of the hard braking event in the range between -7.3 $m/s^2$ and -6.5 $m/s^2$ is higher. The local coordination causes a higher average lane change frequency for HV at low MPRs (i.e., 10\% and 20\%). It starts to decrease after reaching 30\% MPR, whereas in the ad hoc coordination the average lane change frequency maintains the increasing trend. 
The interaction state of the calibrated car-following model for HVs also reveals an interesting pattern. The local coordination yields a higher percentage of ``Free" interaction state than ad hoc coordination. The ``Close-up" state experience a greater reduction in the local coordination strategy, probably due to the shorter following headway, which is also supported by the increase of the ``Follow" state. The percentage of the ``Follow" state is positively correlated to the MPR of CAVs.  With denser traffic flow, it is logical that the more time HVs as a group spend an additional amount of time in the ``Follow" state.

\section{Conclusions}
\label{sect:conclusion}
In this paper, we investigate the two types of clustering strategies for forming CAV platoons. Platoon clustering strategy is of significant importance when it comes to deploying CAV in mixed traffic conditions in the near term. In agreement with most previous studies, CAVs are able to bring benefits to the transportation network with ad hoc coordination, and adapting local coordination can further increase the benefits. In addition, we quantify the impacts of CAV clustering strategies on HVs. The distribution of the hard braking for HVs, when interacting with CAVs, changes significantly with local coordination strategy for platoon formation. In comparison, the distributions for HVs, when interacting with other HVs, does not show significant changes. The average lane change for HVs increases with the presence of CAVs until 30\% MPR is reached. The analysis of the interaction state of the vehicle trajectory data reveals the different impacts of coordination strategy as well. It is shown that the percentage of ``Free'' driving can be increased by the introduction of CAV with local coordination strategy. 

\textcolor{r3}{
Nevertheless, there are some limitations with regard to this study. First, the E-IDM, while being widely adopted, lacks the multi-anticipative car-following characteristic that has been promoted as one of the crucial feature enabled by V2V communication. However, the longitudinal control of CAVs is not the focus of this study, and the local coordination is predominately associated with the lateral control of CAVs. 
Second, the simulation network was calibrated using the field data collected at the northern Virginia, which only represents a subset of the overall driver population. Regional variability could exist among different subsets of drivers.
}
Future research may focus on the following areas. The lateral control is an under-explored area compared to longitudinal control of CAVs. Further investigation of platoon formation in mixed traffic is desired. Currently, there are only a few platoon coordination algorithms, most of which are rule-based. 
Furthermore, the aggressiveness of the lane change for CAVs during platoon formation is also an important aspect. As shown in this paper, the characteristics of the HV traffic could be influenced, and some of CAV maneuvers during clustering could even pose safety concerns for HVs.
Lastly, the comparison among clustering strategies should be further expanded with additional scenarios. 
\textcolor{r3}{Cross validation of different models for CAVs and/or HVs,  as well as the comparison of corresponding results should be conducted in the future study, since the nature of the CAV behavior is still at a developmental stage.}

\appendix
\setcounter{figure}{0}
\setcounter{table}{0}   
\renewcommand{\thetable}{A\arabic{table}}
\label{appendix}
 
\section{\textcolor{r2}{List of Abbreviations}}

\begin{table}[H]
\centering
\caption{\textcolor{r2}{List of Abbreviations}} 
\begin{tabular}{p{1.5in}|p{4.5in}} 
\hline  \hline
\textbf{Abbreviation} & \textbf{Definition} \\ \hline 
ACC & adaptive cruise control \\ \hline 
AH & ad hoc \\ \hline 
AV & automated vehicles  \\ \hline 
CAV & connected and automated vehicles\\ \hline
CACC & cooperative adaptive cruise control \\ \hline 
CAH  & constant acceleration heuristic \\ \hline 
CDF &  cumulative distribution function\\ \hline 
CL & local coordination \\ \hline 
E-IDM & enhanced intelligent driver model \\ \hline 
GP & general purpose \\ \hline 
HOV & high-occupancy vehicle \\ \hline 
HV & human-driven vehicle  \\  \hline  
LMRS  &  Lane-change Model with Relaxation and Synchronization \\ \hline 
MPR &  market penetration rate\\ \hline 
MIXIC &  MICroscopic model for Simulation of Intelligent Cruise control  \\ \hline
MOBIL & minimizing overall braking induced by lane change\\ \hline  
NHCRP & National Highway Cooperative Research Program \\ \hline 
OSDAP & Open Source Application Development Portal  \\ \hline 
SARTRE & Safe Road Trains for the Environment project\\ \hline
SSAM & safety surrogate assessment model \\ \hline 
TTC & time to collision \\ \hline 
VAD &  vehicle awareness device\\ \hline 
VMT & vehicle mile traveled \\ \hline 
VHT & vehicle hour traveled \\ \hline
\end{tabular}
\label{table:abbrv}
\end{table}

\section{\textcolor{re}{CAV Local Coordination Flowchart}}
\label{sect:flowchart}
\begin{figure} [H]
	\centering
	\includegraphics[width=0.9\textwidth]{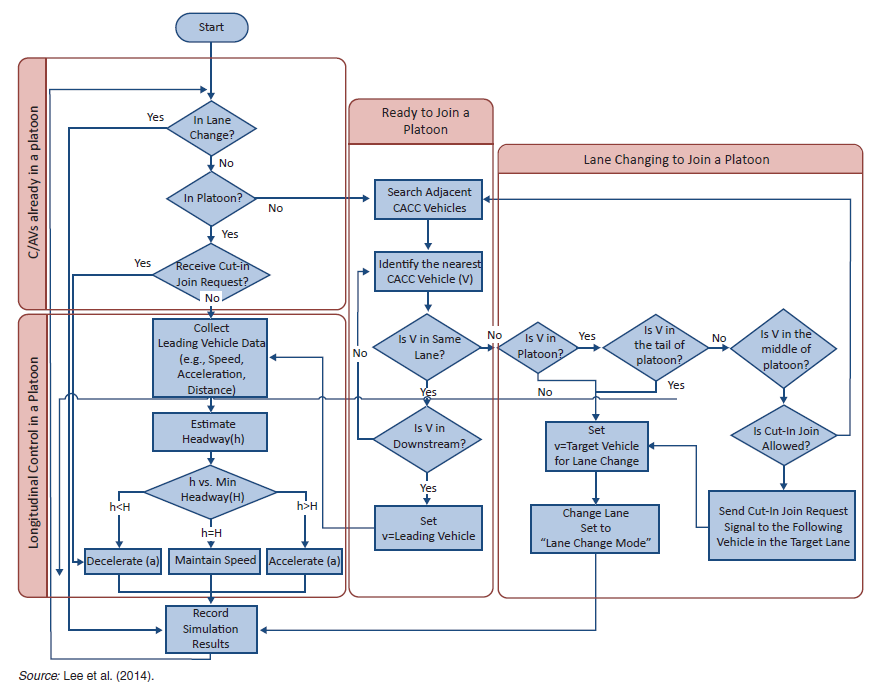}   
	\caption{\textcolor{re}{Flowchart for CACC local coordination} \citep{lee2014mobility, NAP25366}} 
	\label{fig:leeLcAlgo}
\end{figure}

\section{\textcolor{r2}{Simulation Calibration}}
\label{sect:appCalibration}
The data collected from the field for calibration were RTMS (remote traffic microwave sensor), INRIX probe vehicle travel time data, and video camera footage. Each data source is independent and the associated traffic flow metrics are shown in Table \ref{table:dataSource}.

\begin{table}[h]
\centering
\caption{Independent Data Source for Calibration}
\begin{tabular}{cccccc}
\hline  \hline
Data Source &  Volume & Speed & Occupancy & Travel Time & Measurement Type \\ \hline
RTMS data & \checkmark & \checkmark & \checkmark & & point\\
INRIX data & & & & \checkmark & segment\\
Video camera footage & \checkmark  & & & & point\\
\hline 
\end{tabular}
\label{table:dataSource}
\end{table}

\subsection{Origin-destination Demand}

\begin{figure}[H]
\begin{minipage}{.45\linewidth}
\centering
\subfloat[3:30 p.m. - 3:45 p.m.]{\label{main:a}\includegraphics[scale=.38]{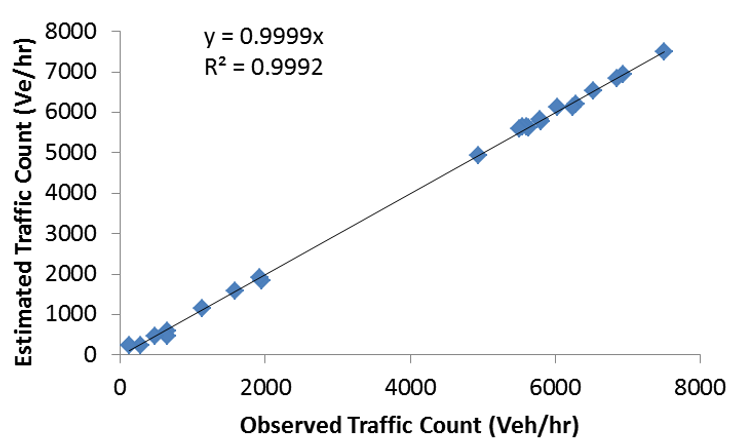}}
\end{minipage}
\begin{minipage}{.45\linewidth}
\centering 
\subfloat[6:15 p.m. - 6:30 p.m.]{\label{mainb}\includegraphics[scale=.38]{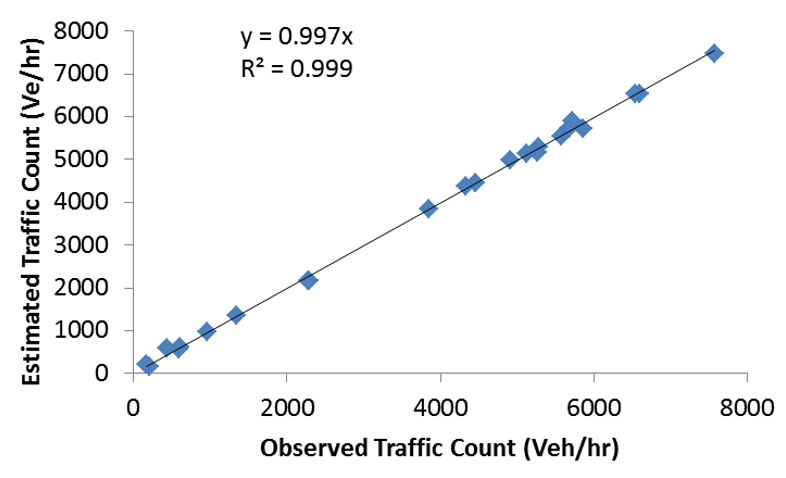}}
\end{minipage}   
\caption{Calibration for OD demand}
\label{fig:calOD}
\end{figure}

The QueensOD \citep{rakha2002queensod}, a software application for estimating time-dependent static O-D, was used to estimate the volumes of the OD pairs that do not have sensor coverage. The results were then compared with field observations at a 15-min interval. Fig. \ref{fig:calOD} displays two of the total 16 intervals for the I-66 Vissim network.

\subsection{Traffic flow characteristics}
\begin{figure}[H]
\begin{minipage}{.45\linewidth}
\centering
\subfloat[RTMS 1 speed-flow diagram]{\label{main:a}\includegraphics[scale=.35]{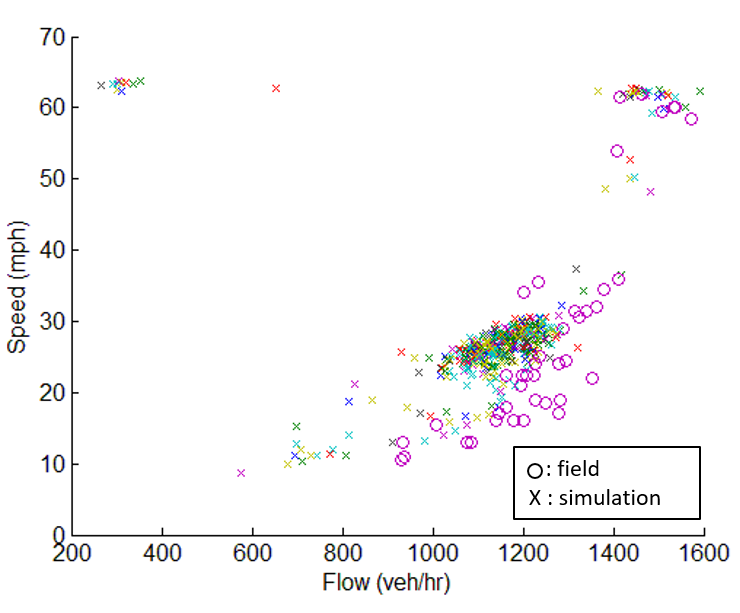}}
\end{minipage}
\begin{minipage}{.45\linewidth}
\centering 
\subfloat[RTMS 2 speed-flow diagram]{\label{mainb}\includegraphics[scale=.35]{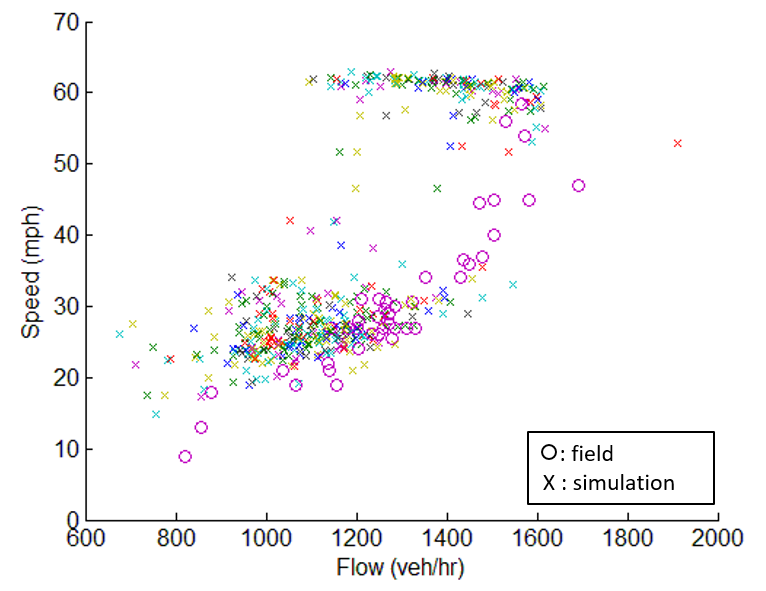}}
\end{minipage}   
\caption{Calibration for traffic flow characteristics}
\label{fig:rtms}
\end{figure}

With the estimated values for all OD pairs, the Latin Hypercube Sampling was adopted to generate hundreds of scenarios for Vissim evaluation in order to narrow down parameters set candidates. Then the selected candidates were fine-tuned by adjusting the driving behaviors in multiple locations to reflect the traffic flow characteristics across multiple segments in the network.  
Figure \ref{fig:rtms} compares the traffic flow characteristics of the two RTMS locations between the field and the simulation at the two RTMS locations. 

The TMC segments shown in Table \ref{table:tmcSeg} were also calibrated. The field travel time from 3:30 p.m. to 6:30 p.m. on Nov. 12, 2014, was used. Note that the daily variance of the travel time was not taken into account in this calibration.

\begin{table}[H]
\centering
\caption{INRIX TMC Segments}
\resizebox{\textwidth}{!}
{
\begin{tabular}{l|l|l|l|l|l|l|l}
\hline \hline
TMC Code & \multicolumn{3}{c}{From (exit, latitude, longitude)} & \multicolumn{3}{c}{To (exit,latitude, longitude)} & Mileage \\ \hline
110+04176 & I-495/EXIT 64 & 38.88308 & -77.2295 & VA-243/NUTLEY ST/EXIT 62 & 38.87883 & -77.2628 & 1.8 \\
110P04176 & I-495/EXIT 64 & 38.87883 & -77.2628 & VA-243/NUTLEY ST/EXIT 62 & 38.87824 & -77.2703 & 0.4 \\
110+04177 & VADEN DR/ EXIT  62 & 38.87824 & -77.2703 & VA-123/EXIT 60 & 38.87741 & -77.2748 & 0.3 \\
110+04178 & VA-243/NUTLEY ST/EXIT 62 & 38.87741 & -77.2748 & VA-123/EXIT 60 & 38.8704 & -77.3006 & 1.4 \\
\hline
\end{tabular}
}
\label{table:tmcSeg}
\end{table}

\begin{figure}[H]
\begin{minipage}{.5\linewidth}
\centering
\subfloat[TMC 110+01476]{\includegraphics[scale=0.35]{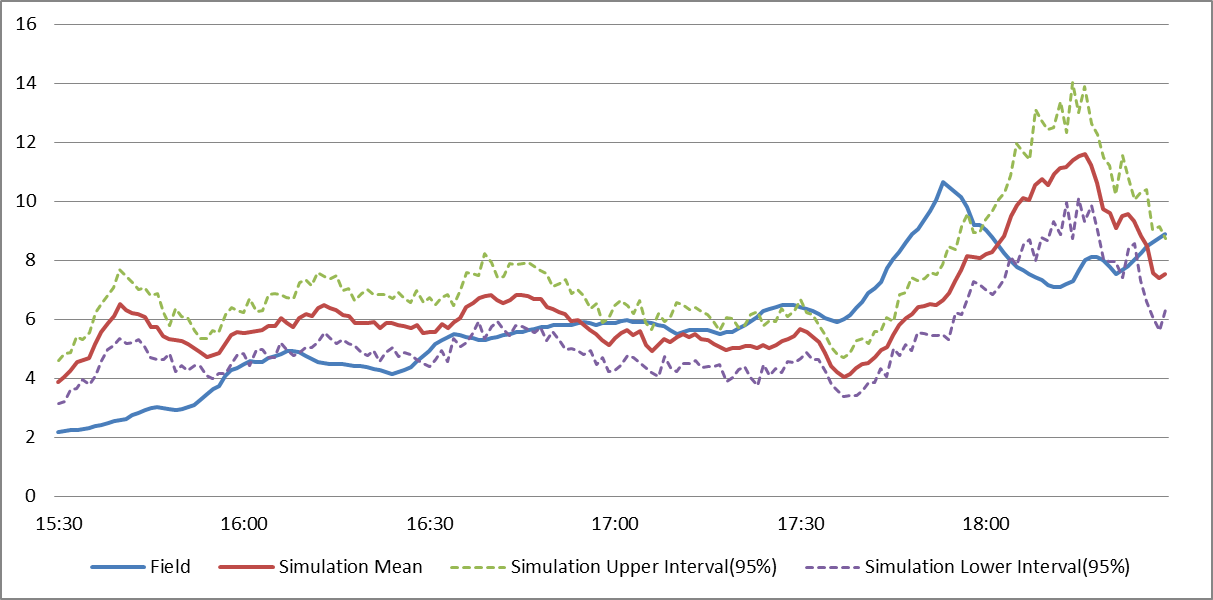}}
\end{minipage}
\begin{minipage}{.5\linewidth}
\centering 
\subfloat[TMC 110P01476]{\includegraphics[scale=.35]{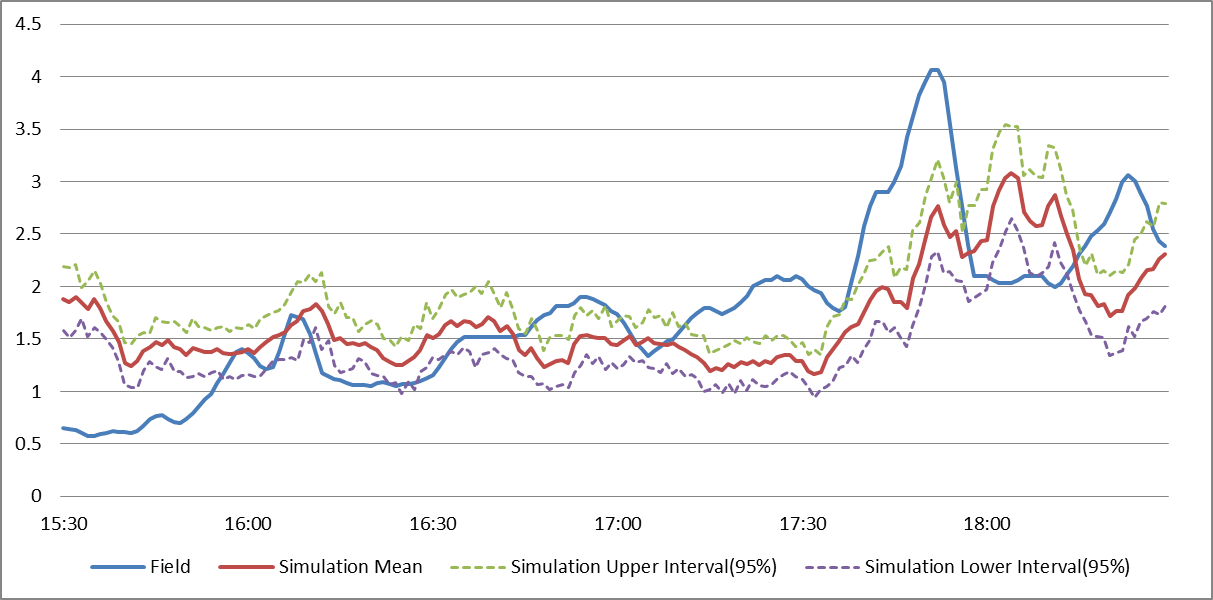}}
\end{minipage}  \par \medskip
\begin{minipage}{.5\linewidth}
\centering 
\subfloat[TMC 110+01477]{\includegraphics[scale=.35]{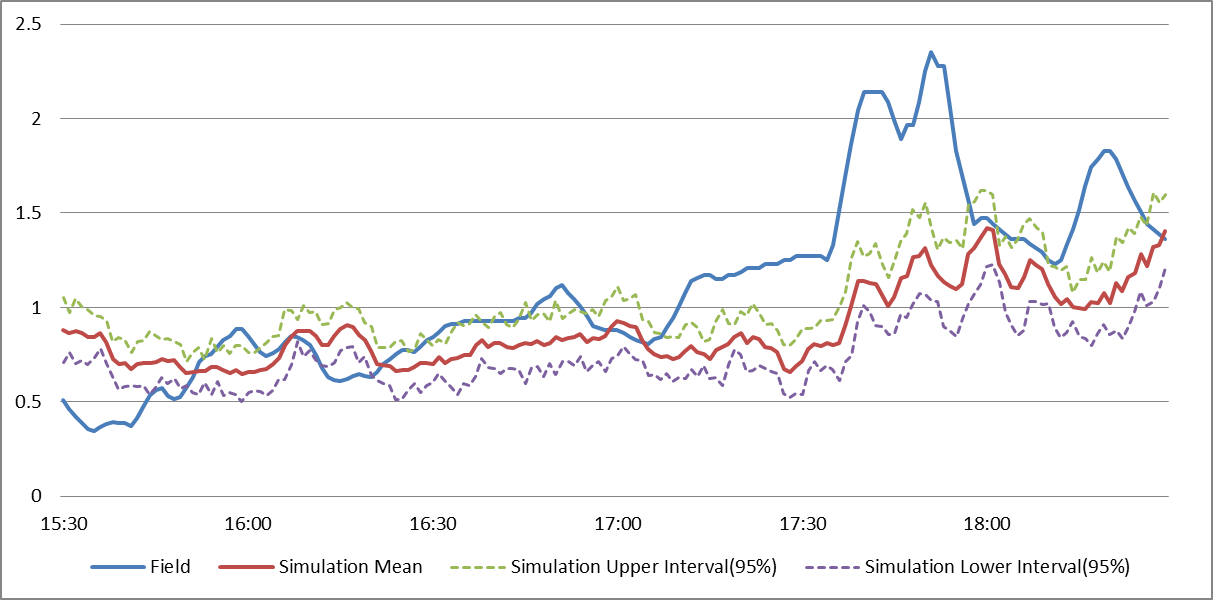}}
\end{minipage}   
\begin{minipage}{.5\linewidth}
\centering 
\subfloat[TMC 110+01478]{\includegraphics[scale=.35]{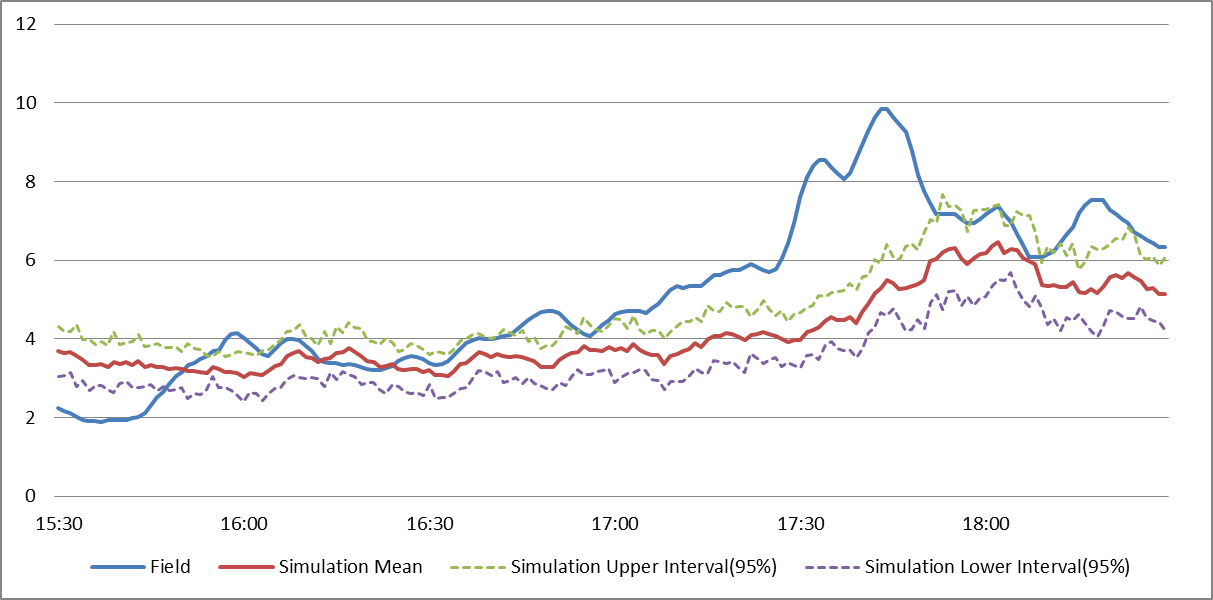}}
\end{minipage}   
  
\caption{Calibration for TMC travel time}
\label{fig:calTmc}
\end{figure}

For the deceleration profile, we used the Vissim default profile for passenger cars, which has the maximum deceleration of -8.5 $m/s^2$. The default profile for maximum deceleration for passenger cars corresponds to those established in the traffic flow model Wiedemann 74. The parameters were measured in Germany, then later, a slight change for shorter time step with a constraint on jerk was adopted \citep{ptv2018ptv}. 
The range, when applicable, of the most influential parameters for the Wiedemann 99 model is shown in Table \ref{table:w99Para}, \textcolor{r3}{which are calibration based on the field data collected in northern Virginia in the United States.} 
\begin{table}[H]
\centering
\caption{Independent Data Source for Calibration}
\begin{tabular}{l|c}
\hline \hline
Wiedemann 99 parameter & Value range \\ \hline
CC0 (standstill distance) & 1.5 m \\
CC1 (following distance) & 0.5 -1.5 s \\
CC2 (longitudinal oscillation) & 3 - 4 m \\
\hline
\end{tabular}
\label{table:w99Para}
\end{table}

\bibliographystyle{elsarticle-harv}
\bibliography{CAV-HV_impact_R3}

\begin{thebibliography}{74}
\expandafter\ifx\csname natexlab\endcsname\relax\def\natexlab#1{#1}\fi
\expandafter\ifx\csname url\endcsname\relax
  \def\url#1{\texttt{#1}}\fi
\expandafter\ifx\csname urlprefix\endcsname\relax\def\urlprefix{URL }\fi

\bibitem[{Aarts and Feddes(2016)}]{aarts2016european}
Aarts, L., Feddes, G., 2016. European truck platooning challenge. In:
  International Symposium on Heavy Vehicle Transport Technology {(HVTT14)},
  Rotorua, New Zealand.

\bibitem[{Amirgholy et~al.(2020)Amirgholy, Shahabi, and
  Gao}]{Amirgholy2020traffic}
Amirgholy, M., Shahabi, M., Gao, H.~O., 2020. Traffic automation and lane
  management for communicant, autonomous, and human-driven vehicles.
  Transportation Research Part C: Emerging Technologies 111, 477 -- 495.
\newline\urlprefix\url{http://www.sciencedirect.com/science/article/pii/S0968090X18315791}

\bibitem[{Arnaout and Arnaout(2014)}]{Arnaout2014}
Arnaout, G.~M., Arnaout, J.-P., 2014. Exploring the effects of cooperative
  adaptive cruise control on highway traffic flow using microscopic traffic
  simulation. Transportation Planning and Technology 37~(2), 186--199.
\newline\urlprefix\url{https://doi.org/10.1080/03081060.2013.870791}

\bibitem[{Bagdadi and V{\'a}rhelyi(2011)}]{bagdadi2011jerky}
Bagdadi, O., V{\'a}rhelyi, A., 2011. Jerky driving—an indicator of accident
  proneness? Accident Analysis \& Prevention 43~(4), 1359--1363.

\bibitem[{Bergh{\"o}fer et~al.(2018)Bergh{\"o}fer, Purucker, Naujoks,
  Wiedemann, and Marberger}]{berghofer2018prediction}
Bergh{\"o}fer, F.~L., Purucker, C., Naujoks, F., Wiedemann, K., Marberger, C.,
  2018. Prediction of take-over time demand in conditionally automated
  driving-results of a real world driving study. In: Proceedings of the Human
  Factors and Ergonomics Society Europe Chapter 2018 Annual Conference. pp.
  69--81.

\bibitem[{{Caltrans Division of Traffic Operations}(2013)}]{Caltrain2013aGuide}
{Caltrans Division of Traffic Operations}, 2013. A guide to using the caltrans
  performance measurement system {(PeMS)} in transportation concept reports.
  Tech. rep.

\bibitem[{Casey and Lund(1992)}]{casey1992changes}
Casey, S.~M., Lund, A.~K., 1992. Changes in speed and speed adaptation
  following increase in national maximum speed limit. Journal of Safety
  Research 23~(3), 135 -- 146.
\newline\urlprefix\url{http://www.sciencedirect.com/science/article/pii/0022437592900163}

\bibitem[{Chan(2016)}]{chan2016sartre}
Chan, E., 2016. {SARTRE} Automated Platooning Vehicles. John Wiley and Sons,
  Ltd, Ch.~10, pp. 137--150.
\newline\urlprefix\url{https://onlinelibrary.wiley.com/doi/abs/10.1002/9781119307785.ch10}

\bibitem[{Chang et~al.(2018)Chang, Fok, et~al.}]{chang2018connected}
Chang, J., Fok, E., et~al., 2018. Connected vehicle pilot positioning and
  timing report: Summary of positioning and timing approaches in {CV} pilot
  sites. Tech. rep., United States. Dept. of Transportation. ITS Joint Program
  Office.

\bibitem[{Eilers et~al.(2015)Eilers, M{\aa}rtensson, Pettersson, Pillado,
  Gallegos, Tobar, Johansson, Ma, Friedrichs, Borojeni,
  et~al.}]{eilers2015companion}
Eilers, S., M{\aa}rtensson, J., Pettersson, H., Pillado, M., Gallegos, D.,
  Tobar, M., Johansson, K.~H., Ma, X., Friedrichs, T., Borojeni, S.~S., et~al.,
  2015. Companion--towards co-operative platoon management of heavy-duty
  vehicles. In: 2015 IEEE 18th International Conference on Intelligent
  Transportation Systems. IEEE, pp. 1267--1273.

\bibitem[{Fairclough et~al.(1997)Fairclough, May, and
  Carter}]{FAIRCLOUGH1997387}
Fairclough, S.~H., May, A.~J., Carter, C., 1997. The effect of time headway
  feedback on following behaviour. Accident Analysis \& Prevention 29~(3), 387
  -- 397.
\newline\urlprefix\url{http://www.sciencedirect.com/science/article/pii/S0001457597000055}

\bibitem[{Fellendorf and Vortisch(2010)}]{fellendorf2010microscopic}
Fellendorf, M., Vortisch, P., 2010. Microscopic traffic flow simulator
  {Vissim}. In: Fundamentals of traffic simulation. Springer, pp. 63--93.

\bibitem[{{FHWA OSDAP}(2015)}]{OSDAP2015}
{FHWA OSDAP}, 2015. Vissim external driver model {(VEDM)} add-on for connected
  automated vehicle ({CAV}) applications now available for download.
\newline\urlprefix\url{https://www.itsforge.net/index.php/vedm-cac-announced}

\bibitem[{Gettman et~al.(2008)Gettman, Pu, Sayed, and
  Shelby}]{gettman2008surrogate}
Gettman, D., Pu, L., Sayed, T., Shelby, S., 2008. Surrogate safety assessment
  model and validation: Final report. Publication No. FHWA-HRT-08 51.

\bibitem[{Ghiasi et~al.(2017)Ghiasi, Hussain, Qian, and Li}]{Ghiasi2017amixed}
Ghiasi, A., Hussain, O., Qian, Z.~S., Li, X., 2017. A mixed traffic capacity
  analysis and lane management model for connected automated vehicles: A markov
  chain method. Transportation Research Part B: Methodological 106, 266 -- 292.
\newline\urlprefix\url{http://www.sciencedirect.com/science/article/pii/S0191261517302278}

\bibitem[{Ghiasi et~al.(2019)Ghiasi, Li, and Ma}]{Ghiasi2019mixed}
Ghiasi, A., Li, X., Ma, J., 2019. A mixed traffic speed harmonization model
  with connected autonomous vehicles. Transportation Research Part C: Emerging
  Technologies 104, 210 -- 233.
\newline\urlprefix\url{http://www.sciencedirect.com/science/article/pii/S0968090X18304169}

\bibitem[{Gipps(1981)}]{Gipps1981A}
Gipps, P., 1981. A behavioural car-following model for computer simulation.
  Transportation Research Part B: Methodological 15~(2), 105 -- 111.
\newline\urlprefix\url{http://www.sciencedirect.com/science/article/pii/0191261581900370}

\bibitem[{Goodman(1954)}]{goodman1954kolmogorov}
Goodman, L.~A., 1954. Kolmogorov-smirnov tests for psychological research.
  Psychological bulletin 51~(2), 160.

\bibitem[{Gouy(2013)}]{gouy2013behavioural}
Gouy, M., 2013. Behavioural adaption of drivers of unequipped vehicles to short
  time headways observed in a vehicle platoon. Ph.D. thesis, The Open
  University.

\bibitem[{Gouy et~al.(2014)Gouy, Wiedemann, Stevens, Brunett, and
  Reed}]{GOUY2014264}
Gouy, M., Wiedemann, K., Stevens, A., Brunett, G., Reed, N., 2014. Driving next
  to automated vehicle platoons: How do short time headways influence
  non-platoon drivers’ longitudinal control? Transportation Research Part F:
  Traffic Psychology and Behaviour 27, 264 -- 273.
\newline\urlprefix\url{http://www.sciencedirect.com/science/article/pii/S1369847814000345}

\bibitem[{Guériau et~al.(2016)Guériau, Billot, Faouzi, Monteil, Armetta, and
  Hassas}]{gueriau2016assess}
Guériau, M., Billot, R., Faouzi, N.-E.~E., Monteil, J., Armetta, F., Hassas,
  S., 2016. How to assess the benefits of connected vehicles? a simulation
  framework for the design of cooperative traffic management strategies.
  Transportation Research Part C: Emerging Technologies 67, 266 -- 279.
\newline\urlprefix\url{http://www.sciencedirect.com/science/article/pii/S0968090X16000462}

\bibitem[{Higgs et~al.(2011)Higgs, Abbas, and Medina}]{higgs2011analysis}
Higgs, B., Abbas, M.~M., Medina, A., 2011. Analysis of the {Wiedemann} car
  following model over different speeds using naturalistic data. In: Procedia
  of RSS Conference. pp. 1--22.

\bibitem[{Hussain et~al.(2016)Hussain, Ghiasi, and Li}]{hussain2016freeway}
Hussain, O., Ghiasi, A., Li, X., 2016. Freeway lane management approach in
  mixed traffic environment with connected autonomous vehicles. arXiv preprint
  arXiv:1609.02946.

\bibitem[{Kesting et~al.(2007)Kesting, Treiber, and Helbing}]{Kesting2007}
Kesting, A., Treiber, M., Helbing, D., 2007. General lane-changing model
  {MOBIL} for car-following models. Transportation Research Record 1999,
  86--94.
\newline\urlprefix\url{http://trrjournalonline.trb.org/doi/10.3141/1999-10}

\bibitem[{Kesting et~al.(2010)Kesting, Treiber, and Helbing}]{Kesting2010}
Kesting, A., Treiber, M., Helbing, D., 2010. {Enhanced intelligent driver model
  to access the impact of driving strategies on traffic capacity}.
  Philosophical Transactions of the Royal Society A: Mathematical, Physical and
  Engineering Sciences 368~(1928), 4585--4605.

\bibitem[{Kesting et~al.(2008)Kesting, Treiber, Schönhof, and
  Helbing}]{Kesting2008adaptive}
Kesting, A., Treiber, M., Schönhof, M., Helbing, D., 2008. Adaptive cruise
  control design for active congestion avoidance. Transportation Research Part
  C: Emerging Technologies 16~(6), 668 -- 683.
\newline\urlprefix\url{http://www.sciencedirect.com/science/article/pii/S0968090X08000028}

\bibitem[{Lank et~al.(2011)Lank, Haberstroh, and Wille}]{Lank2011Interaction}
Lank, C., Haberstroh, M., Wille, M., 2011. Interaction of human, machine, and
  environment in automated driving systems. Transportation Research Record
  2243~(1), 138--145.
\newline\urlprefix\url{https://doi.org/10.3141/2243-16}

\bibitem[{Lee et~al.(2014)Lee, Bared, and Park}]{lee2014mobility}
Lee, J., Bared, J., Park, B., 2014. Mobility impacts of cooperative adaptive
  cruise control ({CACC}) under mixed traffic conditions. In: 93rd Annual
  Meeting of the Transportation Research Board, Washington, DC.

\bibitem[{Lee et~al.(2019)Lee, Jeong, Oh, and Oh}]{Lee2019driving}
Lee, S., Jeong, E., Oh, M., Oh, C., 2019. Driving aggressiveness management
  policy to enhance the performance of mixed traffic conditions in automated
  driving environments. Transportation Research Part A: Policy and Practice
  121, 136 -- 146.
\newline\urlprefix\url{http://www.sciencedirect.com/science/article/pii/S0965856418304063}

\bibitem[{Leidos(2016)}]{STOLT4}
Leidos, 2016. Simulation of evolutionary introduction of cooperative adaptive
  cruise control equipped vehicles into traffic. Tech. rep., Saxton
  Transportation Operations Laboratory.

\bibitem[{Li et~al.(2019)Li, Ma, and Hale}]{Li2019High}
Li, T., Ma, J., Hale, D.~K., 2019. High-occupancy vehicle lanes on the right:
  an alternative design for congestion reduction at freeway merge, diverge, and
  weaving areas. Transportation Letters 0~(0), 1--13.
\newline\urlprefix\url{https://doi.org/10.1080/19427867.2019.1584347}

\bibitem[{Liu et~al.(2018)Liu, Kan, Shladover, Lu, and
  Ferlis}]{liu2018Modeling}
Liu, H., Kan, X.~D., Shladover, S.~E., Lu, X.-Y., Ferlis, R.~E., 2018. Modeling
  impacts of cooperative adaptive cruise control on mixed traffic flow in
  multi-lane freeway facilities. Transportation Research Part C: Emerging
  Technologies 95, 261 -- 279.
\newline\urlprefix\url{http://www.sciencedirect.com/science/article/pii/S0968090X18310313}

\bibitem[{Lu et~al.(2014)Lu, Lee, Chen, Bared, Dailey, and
  Shladover}]{lu2014freeway}
Lu, X.-Y., Lee, J., Chen, D., Bared, J., Dailey, D., Shladover, S.~E., 2014.
  Freeway micro-simulation calibration: Case study using {Aimsun} and {Vissim}
  with detailed field data. In: 93rd Annual Meeting of the Transportation
  Research Board, Washington, DC.

\bibitem[{Ma et~al.(2019)Ma, Hu, Leslie, Zhou, Huang, and Bared}]{Ma2019anEco}
Ma, J., Hu, J., Leslie, E., Zhou, F., Huang, P., Bared, J., 2019. An eco-drive
  experiment on rolling terrains for fuel consumption optimization with
  connected automated vehicles. Transportation Research Part C: Emerging
  Technologies 100, 125 -- 141.
\newline\urlprefix\url{http://www.sciencedirect.com/science/article/pii/S0968090X18300561}

\bibitem[{Milanés and Shladover(2014)}]{MILANES2014285}
Milanés, V., Shladover, S.~E., 2014. Modeling cooperative and autonomous
  adaptive cruise control dynamic responses using experimental data.
  Transportation Research Part C: Emerging Technologies 48, 285 -- 300.
\newline\urlprefix\url{http://www.sciencedirect.com/science/article/pii/S0968090X14002447}

\bibitem[{Naujoks et~al.(2017)Naujoks, Purucker, Wiedemann, Neukum, Wolter, and
  Steiger}]{naujoks2017driving}
Naujoks, F., Purucker, C., Wiedemann, K., Neukum, A., Wolter, S., Steiger, R.,
  2017. Driving performance at lateral system limits during partially automated
  driving. Accident Analysis \& Prevention 108, 147--162.

\bibitem[{Nowakowski et~al.(2010)Nowakowski, O'Connell, Shladover, and
  Cody}]{nowakowski2010cooperative}
Nowakowski, C., O'Connell, J., Shladover, S.~E., Cody, D., 2010. Cooperative
  adaptive cruise control: Driver acceptance of following gap settings less
  than one second. In: Proceedings of the Human Factors and Ergonomics Society
  Annual Meeting. Vol.~54. SAGE Publications Sage CA: Los Angeles, CA, pp.
  2033--2037.

\bibitem[{Nowakowski et~al.(2011)Nowakowski, Shladover, Cody, Bu, O'Connell,
  Spring, Dickey, and Nelson}]{nowakowski2011cooperative}
Nowakowski, C., Shladover, S.~E., Cody, D., Bu, F., O'Connell, J., Spring, J.,
  Dickey, S., Nelson, D., 2011. Cooperative adaptive cruise control: Testing
  drivers' choices of following distances. Tech. rep.

\bibitem[{Papadoulis et~al.(2019)Papadoulis, Quddus, and
  Imprialou}]{Papadoulis2019evaluating}
Papadoulis, A., Quddus, M., Imprialou, M., 2019. Evaluating the safety impact
  of connected and autonomous vehicles on motorways. Accident Analysis \&
  Prevention 124, 12 -- 22.
\newline\urlprefix\url{http://www.sciencedirect.com/science/article/pii/S0001457518306018}

\bibitem[{{PTV Group}(2018)}]{ptv2018ptv}
{PTV Group}, 2018. {PTV Vissim 11 user manual}. Germany: PTV GROUP.

\bibitem[{Qom et~al.(2016)Qom, Xiao, and Hadi}]{qom2016evaluation}
Qom, S.~F., Xiao, Y., Hadi, M., 2016. Evaluation of cooperative adaptive cruise
  control {(CACC)} vehicles on managed lanes utilizing macroscopic and
  mesoscopic simulation. In: Transportation Research Board 95th Annual Meeting.
  No. 16-6384.

\bibitem[{Rakha(2002)}]{rakha2002queensod}
Rakha, H., 2002. Queensod rel. 2.10-user’s guide: Estimating
  origin-destination traffic demands from link flow counts. ed: Michel Van
  Aerde \& Associates Ltd., Blacksburg, VA.

\bibitem[{Reiter(1994)}]{reiter1994empirical}
Reiter, U., 1994. Empirical studies as basis for traffic flow models. In:
  Proceedings of the 2nd International Symposium on Highway Capacity, Vol. 2.

\bibitem[{S.~C.~Calvert and van Lint(2017)}]{Calvert2017will}
S.~C.~Calvert, W. J.~S., van Lint, J. W.~C., 2017. Will automated vehicles
  negatively impact traffic flow. Journal of Advanced Transportation 2017.

\bibitem[{Saha et~al.(2019)Saha, Roy, Sarkar, and Pal}]{Saha2019preferred}
Saha, P., Roy, R., Sarkar, A.~K., Pal, M., 2019. Preferred time headway of
  drivers on two-lane highways with heterogeneous traffic. Transportation
  Letters 11~(4), 200--207.
\newline\urlprefix\url{https://doi.org/10.1080/19427867.2017.1312859}

\bibitem[{Schakel et~al.(2012)Schakel, Knoop, and van Arem}]{Schakel2012}
Schakel, W.~J., Knoop, V.~L., van Arem, B., jan 2012. Integrated lane change
  model with relaxation and synchronization. Transportation Research Record
  2316~(1), 47--57.
\newline\urlprefix\url{https://doi.org/10.3141/2316-06}

\bibitem[{{Schakel} et~al.(2010){Schakel}, {van Arem}, and
  {Netten}}]{Schakel2010effects}
{Schakel}, W.~J., {van Arem}, B., {Netten}, B.~D., Sep. 2010. Effects of
  cooperative adaptive cruise control on traffic flow stability. In: 13th
  International IEEE Conference on Intelligent Transportation Systems. pp.
  759--764.

\bibitem[{Segata et~al.(2012)Segata, Dressler, Lo~Cigno, and
  Gerla}]{segata2012simulation}
Segata, M., Dressler, F., Lo~Cigno, R., Gerla, M., 2012. A simulation tool for
  automated platooning in mixed highway scenarios. In: Proceedings of the 18th
  annual international conference on Mobile computing and networking. ACM, pp.
  389--392.

\bibitem[{Shen and Neyens(2017)}]{shen2017assessing}
Shen, S., Neyens, D.~M., 2017. Assessing drivers' response during automated
  driver support system failures with non-driving tasks. Journal of Safety
  Research 61, 149 -- 155.
\newline\urlprefix\url{http://www.sciencedirect.com/science/article/pii/S0022437517301433}

\bibitem[{Shewmake and Jarvis(2014)}]{shewmake2014hybrid}
Shewmake, S., Jarvis, L., 2014. Hybrid cars and {HOV} lanes. Transportation
  Research Part A: Policy and Practice 67, 304--319.

\bibitem[{Shladover et~al.(2012)Shladover, Su, and Lu}]{shladover2012impacts}
Shladover, S.~E., Su, D., Lu, X.-Y., 2012. Impacts of cooperative adaptive
  cruise control on freeway traffic flow. Transportation Research Record
  2324~(1), 63--70.

\bibitem[{Songchitruksa et~al.(2016)Songchitruksa, Bibeka, Lin, Zhang,
  et~al.}]{songchitruksa2016incorporating}
Songchitruksa, P., Bibeka, A., Lin, L.~I., Zhang, Y., et~al., 2016.
  Incorporating driver behaviors into connected and automated vehicle
  simulation. Tech. rep., Center for Advancing Transportation Leadership and
  Safety (ATLAS Center).

\bibitem[{{Spiliopoulou} et~al.(2017){Spiliopoulou}, {Perraki}, {Papageorgiou},
  and {Roncoli}}]{Spiliopoulou2017Exploitation}
{Spiliopoulou}, A., {Perraki}, G., {Papageorgiou}, M., {Roncoli}, C., June
  2017. Exploitation of acc systems towards improved traffic flow efficiency on
  motorways. In: 2017 5th IEEE International Conference on Models and
  Technologies for Intelligent Transportation Systems (MT-ITS). pp. 37--43.

\bibitem[{Talebpour and Mahmassani(2016)}]{talebpour2016influence}
Talebpour, A., Mahmassani, H.~S., 2016. Influence of connected and autonomous
  vehicles on traffic flow stability and throughput. Transportation Research
  Part C: Emerging Technologies 71, 143--163.

\bibitem[{Talebpour et~al.(2016)Talebpour, Mahmassani, and
  Bustamante}]{Talebpour2016Modeling}
Talebpour, A., Mahmassani, H.~S., Bustamante, F.~E., 2016. Modeling driver
  behavior in a connected environment: Integrated microscopic simulation of
  traffic and mobile wireless telecommunication systems. Transportation
  Research Record 2560~(1), 75--86.
\newline\urlprefix\url{https://doi.org/10.3141/2560-09}

\bibitem[{{Transportation Research Board}(2018)}]{NAP25366}
{Transportation Research Board}, 2018. Dedicating lanes for priority or
  exclusive use by connected and automated vehicles. The National Academies
  Press, Washington, DC.
\newline\urlprefix\url{https://www.nap.edu/catalog/25366}

\bibitem[{Treiber et~al.(2000)Treiber, Hennecke, and Helbing}]{Treiber2000}
Treiber, M., Hennecke, A., Helbing, D., 2000. Congested traffic states in
  empirical observations and microscopic simulations, 1805--1824.

\bibitem[{Treiber and Kesting(2013)}]{treiber2013traffic}
Treiber, M., Kesting, A., 2013. Traffic flow dynamics: Data, models and
  simulation. Springer-Verlag Berlin Heidelberg.

\bibitem[{Van~Arem et~al.(2006)Van~Arem, Van~Driel, and Visser}]{van2006impact}
Van~Arem, B., Van~Driel, C.~J., Visser, R., 2006. The impact of cooperative
  adaptive cruise control on traffic-flow characteristics. IEEE Transactions on
  Intelligent Transportation Systems 7~(4), 429--436.

\bibitem[{van Beinum et~al.(2018)van Beinum, Farah, Wegman, and
  Hoogendoorn}]{vanBeinum2018Driving}
van Beinum, A., Farah, H., Wegman, F., Hoogendoorn, S., 2018. Driving behaviour
  at motorway ramps and weaving segments based on empirical trajectory data.
  Transportation Research Part C: Emerging Technologies 92, 426 -- 441.
\newline\urlprefix\url{http://www.sciencedirect.com/science/article/pii/S0968090X18305928}

\bibitem[{Wang et~al.(2019{\natexlab{a}})Wang, Qin, Wang, and
  Chen}]{Wang2019stability}
Wang, H., Qin, Y., Wang, W., Chen, J., 2019{\natexlab{a}}. Stability of
  {CACC}-manual heterogeneous vehicular flow with partial {CACC} performance
  degrading. Transportmetrica B: Transport Dynamics 7~(1), 788--813.
\newline\urlprefix\url{https://doi.org/10.1080/21680566.2018.1517058}

\bibitem[{Wang et~al.(2019{\natexlab{b}})Wang, van Maarseveen, Happee, Tool,
  and van Arem}]{Wang2019Benefits}
Wang, M., van Maarseveen, S., Happee, R., Tool, O., van Arem, B.,
  2019{\natexlab{b}}. Benefits and risks of truck platooning on freeway
  operations near entrance ramp. Transportation Research Record 2673~(8),
  588--602.
\newline\urlprefix\url{https://doi.org/10.1177/0361198119842821}

\bibitem[{Wang et~al.(2017{\natexlab{a}})Wang, Li, and
  Work}]{wang2017Comparing}
Wang, R., Li, Y., Work, D.~B., 2017{\natexlab{a}}. Comparing traffic state
  estimators for mixed human and automated traffic flows. Transportation
  Research Part C: Emerging Technologies 78, 95 -- 110.
\newline\urlprefix\url{http://www.sciencedirect.com/science/article/pii/S0968090X17300517}

\bibitem[{Wang et~al.(2017{\natexlab{b}})Wang, Wu, Hao, Boriboonsomsin, and
  Barth}]{wang2017developing}
Wang, Z., Wu, G., Hao, P., Boriboonsomsin, K., Barth, M., 2017{\natexlab{b}}.
  Developing a platoon-wide eco-cooperative adaptive cruise control {(CACC)}
  system. In: 2017 IEEE Intelligent Vehicles Symposium (IV). IEEE, pp.
  1256--1261.

\bibitem[{Wiedemann(1974)}]{Wiedemann1974}
Wiedemann, R., 1974. {Simulation des Straenverkehrsflusses}. Ph.D. thesis,
  Karlsruhe.

\bibitem[{Wiedemann(1991)}]{wiedemann1991modelling}
Wiedemann, R., 1991. Modelling of rti-elements on multi-lane roads. In: Drive
  Conference (1991: Brussels, Belgium). Vol.~2.

\bibitem[{Xiao et~al.(2018)Xiao, Wang, Schakel, and van
  Arem}]{Xiao2019unravelling}
Xiao, L., Wang, M., Schakel, W., van Arem, B., 2018. Unravelling effects of
  cooperative adaptive cruise control deactivation on traffic flow
  characteristics at merging bottlenecks. Transportation Research Part C:
  Emerging Technologies 96, 380 -- 397.
\newline\urlprefix\url{http://www.sciencedirect.com/science/article/pii/S0968090X1830528X}

\bibitem[{{Xu} and {Peng}(2019)}]{xu2019design}
{Xu}, S., {Peng}, H., 2019. Design, analysis, and experiments of preview path
  tracking control for autonomous vehicles. IEEE Transactions on Intelligent
  Transportation Systems, 1--11.

\bibitem[{Young(1977)}]{young1977proof}
Young, I.~T., 1977. Proof without prejudice: use of the {Kolmogorov-Smirnov}
  test for the analysis of histograms from flow systems and other sources.
  Journal of Histochemistry \& Cytochemistry 25~(7), 935--941.

\bibitem[{Zhang et~al.(2018)Zhang, Ma, Smith, and Liu}]{zhang2018operational}
Zhang, X., Ma, J., Smith, B., Liu, J., 2018. Operational performance evaluation
  of the managed lane strategy for early deployment of cooperative adaptive
  cruise control. Tech. rep.

\bibitem[{Zhao et~al.(2018)Zhao, Ngoduy, Shepherd, Liu, and
  Papageorgiou}]{zhao2018aplatoon}
Zhao, W., Ngoduy, D., Shepherd, S., Liu, R., Papageorgiou, M., 2018. A platoon
  based cooperative eco-driving model for mixed automated and human-driven
  vehicles at a signalised intersection. Transportation Research Part C:
  Emerging Technologies 95, 802 -- 821.
\newline\urlprefix\url{http://www.sciencedirect.com/science/article/pii/S0968090X18307423}

\bibitem[{Zhong(2018)}]{zhong2018assessing}
Zhong, Z., 2018. Assessing the effectiveness of managed lane strategies for the
  rapid deployment of cooperative adaptive cruise control technology. Ph.D.
  thesis, New Jersey Institute of Technology.

\bibitem[{Zhong et~al.(2017)Zhong, Joyoung, and Zhao}]{Zhong2017a}
Zhong, Z., Joyoung, L., Zhao, L., 2017. {Evaluations of Managed Lane Strategies
  for Arterial Deployment of Cooperative Adaptive Cruise Control}. In: 96th
  Transportation Research Board Annual Meeting. Washington DC, USA.

\bibitem[{Zhong and Lee(2019)}]{zhong2019effectiveness}
Zhong, Z., Lee, J., 2019. The effectiveness of managed lane strategies for the
  near-term deployment of cooperative adaptive cruise control. Transportation
  Research Part A: Policy and Practice 129, 257 -- 270.
\newline\urlprefix\url{http://www.sciencedirect.com/science/article/pii/S0965856419303520}

\end{thebibliography}
\end{document}